 \documentclass[]{revtex4-2}

\usepackage{pstricks,pst-all}

\usepackage{textcomp}
\usepackage{gensymb}
\usepackage{subcaption}
\usepackage{wrapfig}
\usepackage{graphicx}
\usepackage{mathtools}
\usepackage{tabularx}
\usepackage{amssymb}
\usepackage{amsmath, nccmath}
\usepackage{commath}
\usepackage{amsthm}
\usepackage{bm}
\usepackage{lineno}
\makeatletter
\newcommand*{\rom}[1]{\expandafter\@slowromancap\romannumeral #1@}
\makeatother

\usepackage{lipsum}
\makeatletter
\def\ps@pprintTitle{%
 \let\@oddhead\@empty
 \let\@evenhead\@empty
 \def\@oddfoot{}%
 \let\@evenfoot\@oddfoot}
\makeatother

\begin{document}



\title{Casimir energy of $N$ magnetodielectric $\delta$-function plates}



 \author{Venkat Abhignan}
 \address{Department of Physics, National Institute of Technology, Tiruchirapalli - 620015, India}

\begin{abstract}
 To investigate Casimir electromagnetic interaction in $N$ bodies, we implement multiple $\delta$-function plates with electric and magnetic properties. We use their optical properties to study the Casimir energy between the plates by implementing multiple scattering formalism. We initially solve Green's functions for two and three plates configurations to obtain their reflection coefficients. Further, the coefficients are implemented in multiple scattering formalism, and a simple method was obtained to depict energy density distribution in the multiple scattering expansions using diagrammatic loops. The Casimir energy for $N$ bodies depends on multiple scattering parameter $\Delta$; this parameter was distributed into nearest neighbour scattering and next-to-nearest neighbour scattering terms represented by different loops depending on reflection, transmission and propagation distance. In this manner, the Casimir energy density was generalized to $N$ plates by identifying a systematic pattern in the representation of diagrammatic loops. 
\end{abstract}





\maketitle


\section{Introduction}
 Quantum electrodynamic fluctuations in a vacuum confined between two parallel perfectly electrically conducting plates were shown to have a force between them\,\cite{Casimir:1948dh}. This force was later generalized for two dielectric slabs\,\cite{Lifshitz:1956zz}. Even though exact results have been obtained for perfectly conducting boundaries, it is generally accepted that calculating Casimir energies and their corresponding forces is a difficult process for dielectric bodies due to the non-additivity property of the interaction.
 
Casimir energies can be described in terms of reflection coefficients of the configurations, for example, using multiple scattering formalism\,\cite{Milton_2008}. Using the optical Fresnel coefficients, Casimir force for parallel multilayered configurations has been pursued by Toma\ifmmode \check{s}\else \v{s}\fi{}\,\cite{mult3,mult4}. The central idea revolves around using the recurrence relations among Fresnel coefficients of neighbourhood media in an iterative manner and extending the result to $N$ bodies. Primarily, the Casimir force acting on a particular planar region is calculated in studies of all the multilayered systems. However, no closed-form expression was solved for $N$ bodies, to our understanding. Increasing the number of layers leads to cumbersome expressions and becomes challenging when concerned with multilayered dielectric cavities. Beyond this, without using recursive relations, the Casimir energy for $N$ dielectric cavities separated by plasma sheets was recently attempted by Allocca et al.\,\cite{Allocca2022}. They studied Barton's model\,\cite{Barton_2005} for plasma sheets to obtain interaction energy for two, three dielectric cavities, and further inductively, the result was extended to $N$ dielectric cavities. 

In this context, we are interested in studying the Casimir interaction of discrete $N$ plates in a vacuum with magnetodielectric properties, while primarily continuous dielectric bodies were studied previously. We utilise $\delta$-function plates \cite{Prachi2012} to obtain Green's functions and deduce Casimir energy using multiple scattering formalism\,\cite{BALIAN1977,BALIAN1978,Kenneth2006,PRLemig}. The physical relevance of $\delta$-function plates has been discussed previously, and the most interesting aspect is their nontrivial boundary conditions\,\cite{Prachi2012,Milton2013}. Balian and Duplantier were instrumental in developing Green's functions based on multiple scattering formalism\,\cite{BALIAN1977} and computing Casimir energies for perfectly conducting thin conductors\,\cite{BALIAN1978}. Recently, the finite part of Casimir energy was derived using multiple scattering formalism for two disjoint bodies by Kenneth and Klich, implementing Lippmann-Schwinger scattering theory\,\cite{Kenneth2006,Kenneth2008} and by Emig et al.\,\cite{PRLemig}.

 We initially solved for Green's functions for $N=1,2,3$ $\delta$-function plates configurations. We extract the reflection coefficients for these configurations by representing the different regions of the Greens function in a matrix representation. We obtain Casimir energies for $N=2,3,4,5$ $\delta$-function plates using these reflection coefficients and their corresponding multiple scattering parameters ($\Delta_{12\cdots N}$'s) from the multiple scattering formalism. Interpreting these multiple scattering parameters for $N=3,4,5$ $\delta$-function plates in a diagrammatic manner, the pattern for $N$ plates seems to be exhibiting. 

We theorise that Casimir energy $\Delta E_{(12\cdots N)}$ of $N$ plates configuration can be represented using the multiple scattering parameter $\Delta_{12\cdots N}$ such as \begin{equation}
 \frac{\Delta E_{(12\cdots N)}}{A} =  \frac{1}{2} \int_{-\infty}^\infty \frac{d\zeta}{2\pi}
\int \frac{d^2k_\perp}{(2\pi)^2} \Bigg[ \ln \Big[ \Delta_{12\cdots N}^H \Big] + \ln \Big[ \Delta_{12\cdots N}^E \Big] \Bigg],
\end{equation} where the integration is over all frequencies and lateral wavenumbers.
 The parameter $\Delta_{12\cdots N}$ can be distributed into nearest neighbour scattering parameter $\Delta_{ij}$ for all $j=i+1$ ($i\in[1,N-1]$ where $i$ and $j$ are adjacent plates) and next-to-nearest neighbour, next-to-next-to-nearest neighbour, $\cdots$ scattering parameter  $\Delta_{ik}$ for all $k\geq i+2$ ($i\in[1,N-2]$ where $i$ and $k$ are not adjacent plates). These parameters give the different possible ways the paths of propagation contribute to the energy. They can be obtained in terms of the optical properties of the plates, reflection coefficients $r_i$, transmission coefficients $t_i$ and with exponential dependence on the distance between the plates $l_{ij}$ described diagrammatically by loops (Expressions derived in Sec. \rom{5}). For instance, the multiple scattering parameter for $N=3$ plates configuration can be obtained, such as \begin{equation}
\Delta_{123}=\Delta_{12}\Delta_{23}+\Delta_{13} \label{MS123}
\end{equation} and diagrammatic loop distribution of this multiple scattering parameter can be visualized using Fig. \ref{Del123}. Similarly, for $N=4$ plates configuration, the multiple scattering parameter can be obtained as \begin{equation} \Delta_{1234}=\Delta_{12}\Delta_{23}\Delta_{34}+\Delta_{12}\Delta_{24}+\Delta_{13}\Delta_{34}+\Delta_{14} \label{MS1234} \end{equation} and the diagrammatic loop distribution is visualized in Fig. \ref{Del1234}.

 \begin{figure}
    \centering 
\begin{pspicture}(4,4)(0,0)
  \psline[linewidth=0mm]{-}(0,0)(0,4.5)
  \psline[linewidth=0mm]{-}(2,0)(2,4.5)
  \psline[linewidth=0mm]{-}(4,0)(4,4.5)
  \psline[linewidth=0.5mm]{<->}(0,0.5)(2,0.5)
  \rput(1,0.75){\textcolor{black}{ \large $l_{12}$}}
\psline[linewidth=0.5mm]{<->}(2,0.5)(4,0.5)
  \rput(3,0.75){\textcolor{black}{ \large $l_{23}$}}
\psset{arrowscale=1.5}
 \psbezier[ArrowInside=->,showpoints=false,ArrowInsideNo=1](0,3.5)(0,4)(2,4)(2,3.5)
 \psbezier[ArrowInside=->,showpoints=false,ArrowInsideNo=1](2,3.5)(2,3)(0,3)(0,3.5)
 \psbezier[ArrowInside=->,showpoints=false,ArrowInsideNo=1](2,3.5)(2,4)(4,4)(4,3.5)
\psbezier[ArrowInside=->,showpoints=false,ArrowInsideNo=1](4,3.5)(4,3)(2,3)(2,3.5)
 \psbezier[ArrowInside=->,showpoints=false,ArrowInsideNo=1](0,2)(0,2.5)(4,2.5)(4,2)
 \psbezier[ArrowInside=->,showpoints=false,ArrowInsideNo=1](4,2)(4,1.5)(0,1.5)(0,2)
\rput(0,-0.25){\textcolor{black}{ \large $r_1$}}
\rput(2,-0.25){\textcolor{black}{ \large $r_2$}}
\rput(2,-0.65){\textcolor{black}{ \large $t_2$}}
\rput(4,-0.25){\textcolor{black}{ \large $r_3$}}
\rput(2,4.8){\textcolor{black}{ \large $\Delta_{123}$}}
\rput(4.7,3.5){\textcolor{black}{ \large $\Delta_{12} \Delta_{23}$}}
\rput(4.4,2){\textcolor{black}{ \large $\Delta_{13}$}}
\end{pspicture} 
\vspace{6mm}
    \caption{Visualizing $\Delta_{123}$ with optical properties, $r_i$ are reflection coefficients and $t_i$ are transmission coefficients of the plates. $l_{ij}$ is the distance between the plates where $i$ and $j$ are adjacent plates.}
    \label{Del123}
\end{figure}
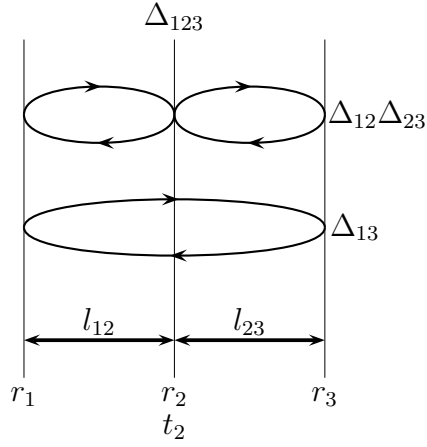
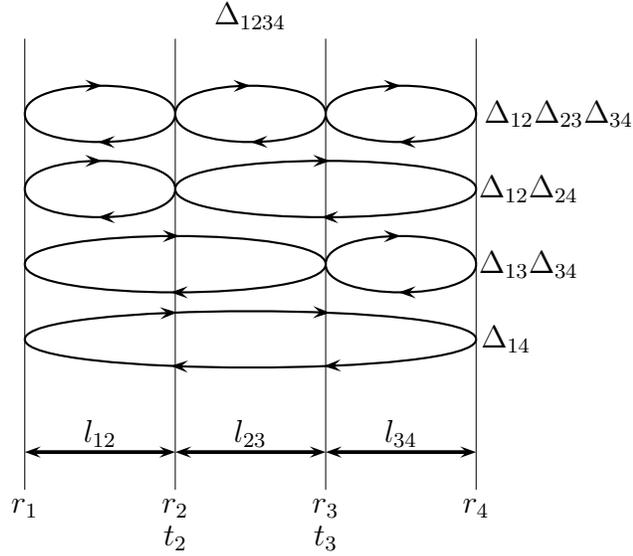
\begin{figure}
    \centering 
\begin{pspicture}(6,6)(0,0)
  \psline[linewidth=0mm]{-}(0,0)(0,6)
  \psline[linewidth=0mm]{-}(2,0)(2,6)
  \psline[linewidth=0mm]{-}(4,0)(4,6)
  \psline[linewidth=0mm]{-}(6,0)(6,6)
  \psline[linewidth=0.5mm]{<->}(0,0.5)(2,0.5)
  \rput(1,0.75){\textcolor{black}{ \large $l_{12}$}}
\psline[linewidth=0.5mm]{<->}(2,0.5)(4,0.5)
  \rput(3,0.75){\textcolor{black}{ \large $l_{23}$}}
  \psline[linewidth=0.5mm]{<->}(4,0.5)(6,0.5)
  \rput(5,0.75){\textcolor{black}{ \large $l_{34}$}}
\psset{arrowscale=1.5}
 \psbezier[ArrowInside=->,showpoints=false,ArrowInsideNo=1](0,5)(0,5.5)(2,5.5)(2,5)
 \psbezier[ArrowInside=->,showpoints=false,ArrowInsideNo=1](2,5)(2,4.5)(0,4.5)(0,5)

\psbezier[ArrowInside=->,showpoints=false,ArrowInsideNo=1](2,5)(2,5.5)(4,5.5)(4,5)
 \psbezier[ArrowInside=->,showpoints=false,ArrowInsideNo=1](4,5)(4,4.5)(2,4.5)(2,5)
 \psbezier[ArrowInside=->,showpoints=false,ArrowInsideNo=1](4,5)(4,5.5)(6,5.5)(6,5)
 \psbezier[ArrowInside=->,showpoints=false,ArrowInsideNo=1](6,5)(6,4.5)(4,4.5)(4,5)

\psbezier[ArrowInside=->,showpoints=false,ArrowInsideNo=1](0,4)(0,4.5)(2,4.5)(2,4)
 \psbezier[ArrowInside=->,showpoints=false,ArrowInsideNo=1](2,4)(2,3.5)(0,3.5)(0,4)

\psbezier[ArrowInside=->,showpoints=false,ArrowInsideNo=1](2,4)(2,4.5)(6,4.5)(6,4)
 \psbezier[ArrowInside=->,showpoints=false,ArrowInsideNo=1](6,4)(6,3.5)(2,3.5)(2,4)

 \psbezier[ArrowInside=->,showpoints=false,ArrowInsideNo=1](0,3)(0,3.5)(4,3.5)(4,3)
 \psbezier[ArrowInside=->,showpoints=false,ArrowInsideNo=1](4,3)(4,2.5)(0,2.5)(0,3)

 \psbezier[ArrowInside=->,showpoints=false,ArrowInsideNo=1](4,3)(4,3.5)(6,3.5)(6,3)
 \psbezier[ArrowInside=->,showpoints=false,ArrowInsideNo=1](6,3)(6,2.5)(4,2.5)(4,3)
 \psbezier[ArrowInside=->,showpoints=false,ArrowInsideNo=2](0,2)(0,2.5)(6,2.5)(6,2)
 \psbezier[ArrowInside=->,showpoints=false,ArrowInsideNo=2](6,2)(6,1.5)(0,1.5)(0,2)

\rput(0,-0.25){\textcolor{black}{ \large $r_1$}}
\rput(2,-0.25){\textcolor{black}{ \large $r_2$}}
\rput(2,-0.65){\textcolor{black}{ \large $t_2$}}
\rput(4,-0.25){\textcolor{black}{ \large $r_3$}}
\rput(4,-0.65){\textcolor{black}{ \large $t_3$}}
\rput(6,-0.25){\textcolor{black}{ \large $r_4$}}
\rput(3,6.3){\textcolor{black}{ \large $\Delta_{1234}$}}
\rput(7.1,5){\textcolor{black}{ \large $\Delta_{12} \Delta_{23}\Delta_{34}$}}
\rput(6.7,4){\textcolor{black}{ \large $\Delta_{12} \Delta_{24}$}}
\rput(6.7,3){\textcolor{black}{ \large $\Delta_{13} \Delta_{34}$}}
\rput(6.4,2){\textcolor{black}{ \large $\Delta_{14}$}}
\end{pspicture}
\vspace{6mm}
    \caption{Visualizing $\Delta_{1234}$ with optical properties, $r_i$ are reflection coefficients and $t_i$ are transmission coefficients of the plates. $l_{ij}$ is the distance between the plates where $i$ and $j$ are adjacent plates.}
     \label{Del1234}
\end{figure}

We compare some of the primary existing works implementing different approaches in the table below:
\begin{center}
\begin{tabularx}{0.98\textwidth}
{ 
  | >{\centering\arraybackslash}X 
  | >{\centering\arraybackslash}X
  | >{\centering\arraybackslash}X 
  | >{\centering\arraybackslash}X | }
 \hline
 &  Toma\ifmmode \check{s}\else \v{s}\fi{}\,\cite{mult3,mult4} & Allocca et al.\,\cite{Allocca2022} & This work \\
 \hline
Configuration  & Multilayered dielectric cavities.  & Multilayered dielectric cavities with plasma sheets as boundaries. & Discrete magnetodielectric $\delta$-function plates in vacuum. \\
\hline
Casimir energy  & Force was derived in the form of Fresnel coefficients obtained from recurrent relations for stacks of layers. & Derived in the form of their so-called generating function (multiple scattering parameter $\Delta_{12\cdots N}$ in our work) obtained from computing ``determinants" related to the transition matrix. & Derived from multiple scattering formalism regarding reflection coefficients obtained from Green's functions. \\
\hline

\end{tabularx}
\end{center}

In Sec.\rom{2}, we introduce the necessary formalism to handle Maxwell's equations in a medium with electric and magnetic properties. Further, electric and magnetic Green's functions are defined, representing the TE and TM modes. In Sec. \rom{3}, Green's functions are solved for a system of one, two and three $\delta$-function plates configurations with electric and magnetic properties in a vacuum. In Sec. \rom{4}, Casimir force is derived between two and three $\delta$-plates using the stress tensor method in terms of their optical properties. In Sec.\rom{5}, we derive the Casimir energy between $N=2,3,4,5$ $\delta$-plates and generalize it to $N$ plates using multiple scattering formalism. Further, we study a particular instance where we consider $N$ plates, which are perfectly dielectric.  
 
\section{Maxwell's equations with TE and TM modes} 
The standard Maxwell's equations for the monochromatic components of electric and magnetic fields $\mathbf{E(r;\omega)}$ and $\mathbf{H(r;\omega)}$ in absence of currents and charges, proportional to $\exp{(-\text{i}\omega t)}$ are 
\begin{subequations}
\begin{eqnarray}
         \nabla \times \mathbf{E}=\text{i}\omega \mathbf{B}, \ \ \ \ \ \ \ \ \ \\ - \nabla \times \mathbf{H}=\text{i}\omega \mathbf{(D+P)}.
         \end{eqnarray} 
\label{3}
\end{subequations} 
These equations imply that $\nabla\cdot\mathbf{B}=0$, $\nabla\cdot(\mathbf{D}+\mathbf{P})=0$ and here $\mathbf{P}$ is the source of external polarization independent of the polarization in the materials due to $\mathbf{E}$ and $\mathbf{H}$. Further here, non-linear responses are neglected in the presence of electric and magnetic materials with boundaries, and macroscopic fields $\mathbf{D}$ and $\mathbf{B}$ respond linearly to $\mathbf{E}$ and $\mathbf{H}$ as \begin{subequations}
\begin{eqnarray}
     \mathbf{D(r;\omega)}=\bm{\varepsilon} \mathbf{(r;\omega)}\cdot\mathbf{E(r;\omega)}, \\
     \mathbf{B(r;\omega)}=\bm{\mu} \mathbf{(r;\omega)}\cdot\mathbf{H(r;\omega)}.
\end{eqnarray}
\end{subequations} $\bm{\varepsilon}$ is dielectric permittivity  and $\bm{\mu}$ magnetic permeability of the materials. The Maxwell's equations decouple and can be combined to obtain second order differential equations for electric field \begin{equation}
\left[ - {\bm\varepsilon}({\bf r};\omega)+ \frac{1}{\omega^2} {\bm\nabla} \times {\bm\mu ({\bf r};\omega)}^{-1} \cdot {\bm\nabla}
\times  \right] \cdot {\bf E}({\bf r},\omega)
= {\bf P}({\bf r},\omega),
\label{ddE=P}
\end{equation} and magnetic field in a similar manner. These correspond to transverse electric (TE) and transverse magnetic (TM) modes rewritten in the form of Green's dyadics $\mathbf{\Gamma}$ and $\mathbf{\Phi}$\,\cite{SCHWINGER1978,Prachi2012} such as \begin{equation}
\left[ - {\bm\varepsilon}({\bf r};\omega)+ \frac{1}{\omega^2} {\bm\nabla} \times {\bm\mu ({\bf r};\omega)}^{-1} \cdot {\bm\nabla}
\times  \right] \cdot 
{\bm\Gamma}({\bf r},{\bf r}^\prime;\omega)
= {\bf 1} \delta^{(3)}({\bf r}-{\bf r}^\prime),
\label{Gd-deq} \end{equation} by correlating the field and source as \begin{equation}
          \mathbf{E(r;\omega)}=\int \hbox{d}^3 r' \mathbf{\Gamma(r,r',\omega)}\cdot\mathbf{P(r',\omega)}. \label{20}
      \end{equation} Green's dyadics $\mathbf{\Gamma}$ and $\mathbf{\Phi}$ can be expressed in form of scalar Green's function $g^E$ and $g^H$, respectively. The scalar components $\gamma^{i j}$ of dyadic $\bm \gamma$ in Fourier space can be derived and represented as a matrix such as \begin{equation}
  \bm \gamma(z,z';\mathbf{k_{\perp}},\omega) = \left[ \begin{array}{ccc}
\frac{1}{\varepsilon^\perp(z)} \frac{\partial}{\partial z}
\frac{1}{\varepsilon^\perp(z^\prime)} \frac{\partial}{\partial z^\prime}
 g^H(z,z^\prime) & 0 &
\frac{1}{\varepsilon^\perp(z)} \frac{\partial}{\partial z}
\frac{ik_\perp}{\varepsilon^{||}(z^\prime)} g^H(z,z^\prime) \\[2mm]
0 & \omega^2 g^E(z,z^\prime) & 0 \\[2mm]
-\frac{ik_\perp}{\varepsilon^{||}(z)} \frac{1}{\varepsilon^\perp(z^\prime)} 
\frac{\partial}{\partial z^\prime} g^H(z,z^\prime) & 0 &
-\frac{ik_\perp}{\varepsilon^{||}(z)}
\frac{ik_\perp}{\varepsilon^{||}(z^\prime)} g^H(z,z^\prime)
\end{array} \right]
-\delta(z-z^\prime)
\left[ \begin{array}{llr}
\frac{1}{\varepsilon^\perp(z)} \hspace{2mm}& 0 \hspace{2mm} & 0 \\
0 & 0 & 0 \\ 0 & 0 & \frac{1}{\varepsilon^{||}(z)} 
\end{array} \right], \label{30} 
\end{equation} from using a two-point correlation function \begin{equation}
{\bm\Gamma}({\bf r},{\bf r}^\prime;\omega)
= \int \frac{d^2k_\perp}{(2\pi)^2} 
\,e^{i{\bf k}_\perp \cdot ({\bf r}-{\bf r}^\prime)_\perp} 
{\bm\gamma}(z,z^\prime;{\bf k}_\perp,\omega),
\end{equation} where $\bm{\varepsilon} (z) = \varepsilon^\perp (z)\, {\bf 1}_\perp 
+ \varepsilon^{||}(z) \, \hat{\bf z} \,\hat{\bf z}$ and
$\bm{\mu} (z) = \mu^\perp (z)\, {\bf 1}_\perp 
+ \mu^{||} (z) \,\hat{\bf z} \,\hat{\bf z}.$ Electric Green's function $g^E$ referring to TE mode and $g^H$ magnetic Green's function referring to TM mode can be defined to satisfy the differential equations such as 
      \begin{equation}
          \left[ - \frac{\partial}{\partial z} \frac{1}{\mu^\perp(z)}
\frac{\partial}{\partial z} + \frac{k_\perp^2}{\mu^{||}(z)} 
-\omega^2 \varepsilon^\perp(z) \right] g^E(z,z^\prime) = \delta(z-z^\prime) \label{18}
      \end{equation} and \begin{equation}
         \left[ - \frac{\partial}{\partial z} \frac{1}{\varepsilon^\perp(z)}
\frac{\partial}{\partial z} + \frac{k_\perp^2}{\varepsilon^{||}(z)} 
-\omega^2 \mu^\perp(z) \right] g^H(z,z^\prime) = \delta(z-z^\prime), \label{19}
      \end{equation} respectively. As a direct consequence of Eq. (\ref{20}), the correlations can be interpreted as\,\cite{Brevik2018} \begin{equation}
          \frac{\delta {\bf E}({\bf r};\omega)}{\delta {\bf P}({\bf r}^\prime;\omega)} = {\bm\Gamma}({\bf r},{\bf r}^\prime;\omega).
      \end{equation} The electric and magnetic field at two distinct points in space can also be correlated using the Green's dyadic in form of Green's functions, and these correlations are expressed as \begin{subequations}
\begin{align}
\frac{1}{\tau}
\langle {\bf E}({\bf r};-\omega) {\bf E}({\bf r}^\prime;\omega) \rangle
&= \frac{1}{\hbox{i}} {\bm\Gamma}({\bf r},{\bf r}^\prime;\omega), \label{EE=iG} \\
\frac{1}{\tau}
\langle {\bf H}({\bf r};-\omega) {\bf H}({\bf r}^\prime;\omega) \rangle
&= \frac{1}{\hbox{i}} {\bm\Gamma}({\bf r},{\bf r}^\prime;\omega)
\Big|_{E\leftrightarrow H, \varepsilon \leftrightarrow \mu}, 
\end{align}
\label{cor-fields-GD}
\end{subequations}
where $\tau$ is the system's average observed time. 
\section{$\delta$-function plates}
In our analysis the interest is in thin $\delta$-function plates\,\cite{Prachi2012,SHAJESH2017} with electric and magnetic properties
\begin{subequations}
\begin{eqnarray}
    \bm{\varepsilon}(z)=\mathbf{1}+\boldsymbol{\lambda}_{ei}(\omega)\delta(z-a_i), \\
    \bm{\mu}(z)=\mathbf{1}+\boldsymbol{\lambda}_{gi}(\omega)\delta(z-a_i),
\end{eqnarray}
\label{32}
\end{subequations}  at positions $z=a_i$ in vacuum up to an infinite extent in $x-y$ axis. Since we consider disjoint objects the latter term in dyadic $\bm \gamma$ in Eq. (\ref{30}) containing $\delta(z-z')$ can be neglected in this case, since it does not contribute. The material properties of $\delta$-function plates, with planar symmetry (homogeneous and isotropic on the plane) are described by the matrix \begin{equation}
  \boldsymbol{\lambda}(\omega) = \left[
\begin{array}{ccc}
 \lambda^{\perp}(\omega) & 0 & 0 \\
 0 & \lambda^{\perp}(\omega) & 0 \\
 0 & 0 & \lambda^{\parallel}(\omega) \\
\end{array}
\right].
\end{equation}
The boundary conditions satisfying the planar interface at $z=a_i$ are derived from implementing an Amperian loop integral and a Gaussian surface integral across the plate. Additional nontrivial contributions to the electric field boundary conditions are obtained due to $\delta$-functions such as \begin{subequations}
\begin{eqnarray}
 \left. E_2(\mathbf{k_{\perp}},z;\omega)\right|^{z=a+\delta}_{z=a-\delta}&=-\text{i}\omega\lambda^{\perp}_{g i}H_1(\mathbf{k_{\perp}},a_i), \\
 \left. E_1(\mathbf{k_{\perp}},z;\omega)\right|^{z=a+\delta}_{z=a-\delta}&=\text{i}\omega\lambda^{\perp}_{g i}H_2(\mathbf{k_{\perp}},a_i),
\\
\left. E_3(\mathbf{k_{\perp}},z;\omega)\right|^{z=a+\delta}_{z=a-\delta}&=-\text{i}k_{\perp}\lambda^{\perp}_{e i} E_1(\mathbf{k_{\perp}},a_i).
\end{eqnarray}
\label{34}
\end{subequations} Similarly, the magnetic field boundary conditions are derived such as 
\begin{subequations}
\begin{eqnarray}
\left. H_2(\mathbf{k_{\perp}},z;\omega)\right|^{z=a+\delta}_{z=a-\delta}&=\text{i}\omega\lambda^{\perp}_{e i}E_1(\mathbf{k_{\perp}},a_i), \\
 \left. H_1(\mathbf{k_{\perp}},z;\omega)\right|^{z=a+\delta}_{z=a-\delta}&=-\text{i}\omega\lambda^{\perp}_{e i}E_2(\mathbf{k_{\perp}},a_i),
\\
\left. H_3(\mathbf{k_{\perp}},z;\omega)\right|^{z=a+\delta}_{z=a-\delta}&=-\text{i}k_{\perp}\lambda^{\perp}_{g i} H_1(\mathbf{k_{\perp}},a_i).
\end{eqnarray}
\label{35}
\end{subequations} Setting $\lambda=0$ would reduce these equations to the traditional well-known boundary conditions for electric and magnetic field in vacuum. The presence of $\delta$-plates with magnetoelectric properties modifies these boundary conditions on the interface making them nontrivial. 
And $\delta$-function plates have boundary constraints \begin{subequations}
\begin{eqnarray}
    \lambda^{\parallel}_{e i}E_3(\mathbf{k_{\perp}},a_i;\omega)=0,\\
    \lambda^{\parallel}_{g i}H_3(\mathbf{k_{\perp}},a_i;\omega)=0.
\end{eqnarray}
\end{subequations}
These reveal that $\lambda^{\parallel}_{e i}=0$ and $\lambda^{\parallel}_{g i}=0$ unless $E_3(\mathbf{k_{\perp}},a_i;\omega)=0$ and $H_3(\mathbf{k_{\perp}},a_i;\omega)=0$ at $z=a_i$. These boundary conditions on electric and magnetic field govern the boundary conditions on TE mode and TM mode from Eq.(\ref{18}) and (\ref{19}), which in turn dictate the boundary conditions on the electric Green's function $g^E$ and magnetic Green's function $g^H$. The boundary conditions on $g^E$ are
\begin{subequations}
\begin{eqnarray}
g^E(z,z^\prime) \Big|^{z=a_i+\delta}_{z=a_i-\delta}
&=& \frac{\lambda^\perp_{g i}}{2}
\left[ \left\{ \frac{\partial}{\partial z}
g^E(z,z^\prime) \right\}_{z=a_i+\delta} 
+ \left\{\frac{\partial}{\partial z}
g^E(z,z^\prime) \right\}_{z=a_i-\delta} \right], \\
 \frac{\partial}{\partial z} g^E(z,z^\prime)\bigg|^{z=a_i+\delta}_{z=a_i-\delta}
&=& \zeta^2 \frac{\lambda^\perp_{e i}}{2}
\big[ g^E(a_i+\delta,z^\prime) + g^E(a_i-\delta,z^\prime) \big].
\end{eqnarray}
\label{37}
\end{subequations}
Similarly, the boundary conditions on $g^H$ are
\begin{subequations}
\begin{eqnarray}
g^H(z,z^\prime) \Big|^{z=a_i+\delta}_{z=a_i-\delta}
&=& \frac{\lambda^\perp_{e i}}{2}
\left[ \left\{  \frac{\partial}{\partial z}
g^H(z,z^\prime) \right\}_{z=a_i+\delta} 
+ \left\{  \frac{\partial}{\partial z}
g^H(z,z^\prime) \right\}_{z=a_i-\delta} \right], \\
  \frac{\partial}{\partial z}
g^H(z,z^\prime) \bigg|^{z=a_i+\delta}_{z=a_i-\delta}
&=& \zeta^2 \frac{\lambda^\perp_{g i}}{2}
\big[ g^H(a_i+\delta,z^\prime) + g^H(a_i-\delta,z^\prime) \big],
\end{eqnarray}
\label{38}
\end{subequations} in vacuum where $\omega \rightarrow \text{i}\zeta$ switching to imaginary frequencies using Euclidean rotation. $\zeta$ is the Matsubara frequency, useful in finite temperature calculations.  

 \begin{figure}[!ht]
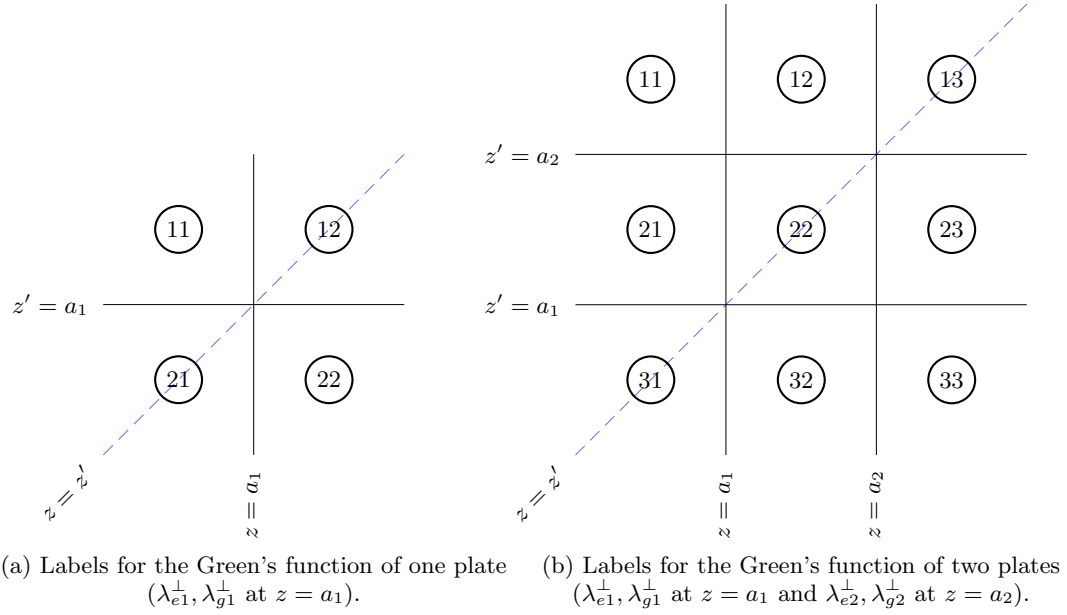

 \centering
 \begin{subfigure}{0.4\textwidth}
 \pspicture(0,0)(4,-4.5)

\psline[linecolor=black,linewidth=0mm]{-}(0,-2)(4,-2) 
\psline[linecolor=black,linewidth=0mm]{-}(2,-0)(2,-4) 
\psline[linecolor=blue,linewidth=0mm,linestyle=dashed,
strokeopacity=0.6]{-}(0,-4)(4,-0) 
\rput[r]{45}(-0.2,-4.2){$z=z^\prime$}
\rput(1,-1){\pscirclebox{11}}
\rput(1,-3){\pscirclebox{21}}
\rput(3,-3){\pscirclebox{22}}
\rput(3,-1){\pscirclebox{12}}
\rput[r]{90}(2.0,-4.2){$z=a_1$}
\rput[r]{0}(-0.2,-2.0){$z^\prime=a_1$}
\endpspicture
\vspace{6mm}
 \caption{Labels for the Green's function of one  plate ($\lambda^{\perp}_{e 1}, \lambda^{\perp}_{g 1}$ at $z=a_1$).}
 \label{1a}
 \end{subfigure}
 \begin{subfigure}{0.4\textwidth}
\pspicture(0,0)(6,-6.5)

\psline[linecolor=black,linewidth=0mm]{-}(0,-2)(6,-2) 
\psline[linecolor=black,linewidth=0mm]{-}(0,-4)(6,-4) 
\psline[linecolor=black,linewidth=0mm]{-}(2,-0)(2,-6) 
\psline[linecolor=black,linewidth=0mm]{-}(4,-0)(4,-6) 
\psline[linecolor=blue,linewidth=0mm,linestyle=dashed,
strokeopacity=0.6]{-}(0,-6)(6,-0) 
\rput(5,-5){\pscirclebox{33}}
\rput(1,-5){\pscirclebox{31}}
\rput(1,-1){\pscirclebox{11}}
\rput(5,-1){\pscirclebox{13}}
\rput(3,-5){\pscirclebox{32}}
\rput(1,-3){\pscirclebox{21}}
\rput(3,-3){\pscirclebox{22}}
\rput(5,-3){\pscirclebox{23}}
\rput(3,-1){\pscirclebox{12}}
\rput[r]{90}(2.0,-6.2){$z=a_1$}
\rput[r]{90}(4.0,-6.2){$z=a_2$}
\rput[r]{0}(-0.2,-2.0){$z^\prime=a_2$}
\rput[r]{0}(-0.2,-4.0){$z^\prime=a_1$}
\rput[r]{45}(-0.2,-6.2){$z=z^\prime$}
\endpspicture
\vspace{6mm}
 \caption{Labels for the Green's function of two plates ($\lambda^{\perp}_{e 1}, \lambda^{\perp}_{g 1}$ at $z=a_1$ and $\lambda^{\perp}_{e 2}, \lambda^{\perp}_{g 2}$ at $z=a_2$).}
 \label{1b}
 \end{subfigure}
 \caption{ Different labels for the regions in the $z-z'$ space for the Green's functions of thin magnetodielectric plates.}\label{Fig1}
 \end{figure}

\begin{figure}[htp]
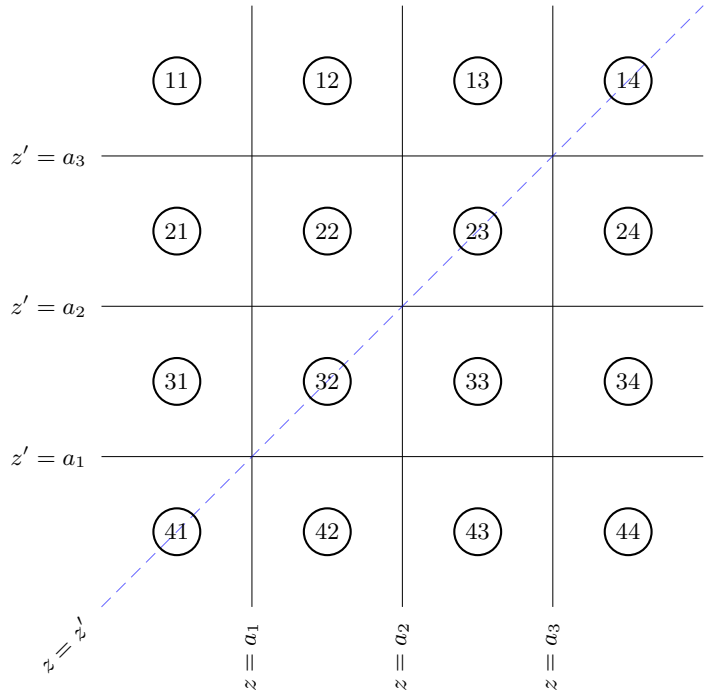

     \begin{center}
      \pspicture(0,0)(8,-8.5)
\psline[linecolor=black,linewidth=0mm]{-}(0,-2)(8,-2) 
\psline[linecolor=black,linewidth=0mm]{-}(0,-4)(8,-4) 
\psline[linecolor=black,linewidth=0mm]{-}(0,-6)(8,-6) 
\psline[linecolor=black,linewidth=0mm]{-}(2,-0)(2,-8) 
\psline[linecolor=black,linewidth=0mm]{-}(4,-0)(4,-8) 
\psline[linecolor=black,linewidth=0mm]{-}(6,-0)(6,-8) 
\psline[linecolor=blue,linewidth=0mm,linestyle=dashed,
strokeopacity=0.6]{-}(0,-8)(8,-0) 
\rput(5,-5){\pscirclebox{33}}
\rput(1,-5){\pscirclebox{31}}
\rput(1,-1){\pscirclebox{11}}
\rput(5,-1){\pscirclebox{13}}
\rput(3,-7){\pscirclebox{42}}
\rput(7,-3){\pscirclebox{24}}
\rput(1,-7){\pscirclebox{41}}
\rput(3,-5){\pscirclebox{32}}
\rput(1,-3){\pscirclebox{21}}
\rput(3,-3){\pscirclebox{22}}
\rput(5,-3){\pscirclebox{23}}
\rput(3,-1){\pscirclebox{12}}
\rput(7,-1){\pscirclebox{14}}
\rput(5,-7){\pscirclebox{43}}
\rput(7,-7){\pscirclebox{44}}
\rput(7,-5){\pscirclebox{34}}
\rput[r]{90}(2.0,-8.2){$z=a_1$}
\rput[r]{90}(4.0,-8.2){$z=a_2$}
\rput[r]{90}(6.0,-8.2){$z=a_3$}
\rput[r]{0}(-0.2,-2.0){$z^\prime=a_3$}
\rput[r]{0}(-0.2,-4.0){$z^\prime=a_2$}
\rput[r]{0}(-0.2,-6.0){$z^\prime=a_1$}
\rput[r]{45}(-0.2,-8.2){$z=z^\prime$}
\endpspicture
\vspace{6mm}
     \caption{Labels for the Green's function of three plates ($\lambda^{\perp}_{e 1}, \lambda^{\perp}_{g 1}$ at $z=a_1$, $\lambda^{\perp}_{e 2}, \lambda^{\perp}_{g 2}$ at $z=a_2$ and $\lambda^{\perp}_{e 3}, \lambda^{\perp}_{g 3}$ at $z=a_3$).}
    \label{Fig2}
    \end{center}
 \end{figure} 

The magnetic Green's functions $g^{1,H}$, $g^{2,H}$ and $g^{3,H}$ are solved using its differential Eq.(\ref{19}) and boundary conditions in Eq.(\ref{38}) for the configuration of one plate with optical properties  $\lambda^{\perp}_{e 1}, \lambda^{\perp}_{g 1}$ at $z=a_1$, two plates with optical properties  $\lambda^{\perp}_{e 1}, \lambda^{\perp}_{g 1}$ at $z=a_1$, $\lambda^{\perp}_{e 2}, \lambda^{\perp}_{g 2}$ at $z=a_2$ and three plates with optical properties  $\lambda^{\perp}_{e 1}, \lambda^{\perp}_{g 1}$ at $z=a_1$, $\lambda^{\perp}_{e 2}, \lambda^{\perp}_{g 2}$ at $z=a_2$, $\lambda^{\perp}_{e 3}, \lambda^{\perp}_{g 3}$ at $z=a_3$, respectively. Illustratively, the solution for $g^H$ for these configurations in different regions of $z-z'$ space is shown in Fig. (\ref{1a}), (\ref{1b}) and (\ref{Fig2}). The solution for $g^{1,H}$ of one plate with the subscript label in Fig. (\ref{1a}) representing the different regions in $z-z'$ space is 
  \begin{subequations}
  \begin{align}
  g^{1,H}_{ \rput(0.1,0){\scriptstyle{11}}
  \pscircle[linewidth=0.2pt](0.1,0.01){0.15} }
  (z,z^\prime) &= \frac{t_1^H}{2 \kappa} \hbox{e}^{-\kappa(a_1-z)} \hbox{e}^{-\kappa(z'-a_1)},
  && z<a_1<z^\prime \\
  g^{1,H}_{ \rput(0.1,0){\scriptstyle{12}}
  \pscircle[linewidth=0.2pt](0.1,0.01){0.15} }
  (z,z^\prime) &= \frac{1}{2 \kappa} \hbox{e}^{-\kappa|z-z'|}+\frac{r_1^H}{2 \kappa} \hbox{e}^{-\kappa(z-a_1)} \hbox{e}^{-\kappa(z'-a_1)},
  && z,z^\prime> a_1
  \end{align}
  \end{subequations}

  and 
 
  \begin{subequations}
  \begin{align}
  g^{1,H}_{ \rput(0.1,0){\scriptstyle{21}}
  \pscircle[linewidth=0.2pt](0.1,0.01){0.15} }
  (z,z^\prime) &= \frac{1}{2 \kappa} \hbox{e}^{-\kappa|z-z'|}+\frac{r_1^H}{2 \kappa} \hbox{e}^{-\kappa(a_1-z)} \hbox{e}^{-\kappa(a_1-z')},
  && z,z^\prime< a_1, \\
  g^{1,H}_{ \rput(0.1,0){\scriptstyle{22}}
  \pscircle[linewidth=0.2pt](0.1,0.01){0.15} }
  (z,z^\prime) &= \frac{t_1^H}{2 \kappa} \hbox{e}^{-\kappa(z-a_1)} \hbox{e}^{-\kappa(a_1-z')}, 
  && z^\prime<a_1<z.
  \end{align}
  \end{subequations}
Here, the reflection and transmission coefficients at plate are \begin{align}
        r_i^H= -\frac{\lambda^{\perp}_{g i} \zeta^2}{\lambda^{\perp}_{g i} \zeta^2+2 \kappa}+\frac{\lambda^{\perp}_{e i} \kappa}{\lambda^{\perp}_{e i} \kappa +2 }, \ t_i^H=1-\frac{\lambda^{\perp}_{g i} \zeta^2}{\lambda^{\perp}_{g i} \zeta^2+2 \kappa}-\frac{\lambda^{\perp}_{e i} \kappa}{\lambda^{\perp}_{e i} \kappa +2 }
        \label{coeff}
         \end{align} for $i=1$ and $\kappa=\sqrt{k_{\perp}^2+\zeta^2}$. Perhaps, a better representation of this Green's function $g^{1,H}$ is from construction of matrices $A$, $B^H$ and $C$ in relation with regions in Fig. 1a such as \begin{equation}
    A = \left[\begin{array}{c c}
 \hbox{e}^{-\kappa(z'-a_1)} &  \hbox{e}^{-\kappa(a_1-z')} \end{array}\right], B^H = \left[\begin{array}{c c} t_1^H & r_1^H \\ r_1^H & t_1^H
\end{array}\right]\,\hbox{and}\, C = \left[\begin{array}{c c}
\hbox{e}^{-\kappa(a_1-z)} \\ \hbox{e}^{-\kappa(z-a_1)} \end{array}\right]. \label{mat}
\end{equation}
 $B^H_{ij}$ represents the element of a matrix $B^H$ in $i^{th}$ row, $j^{th}$ column and the regions in Green's function can be obtained from 
 \begin{equation}
	 g^{N,H}_{ \rput(0.1,0){\scriptstyle{ij}} \pscircle[linewidth=0.2pt](0.1,0.01){0.15} } =
	\begin{cases}
		\frac{1}{2 \kappa} \hbox{e}^{-\kappa|z-z'|}+\frac{1}{2 \kappa} A_{i} B^H_{ij} C_{j}& 
		\text{if}\hspace{2mm} i+j-2=N,\\
		\frac{1}{2 \kappa} A_{i} B^H_{ij} C_{j}&  \text{if}\hspace{2mm} i+j-2\neq N,
	\end{cases}
	\label{matrix}
\end{equation} for $i=1,\cdots,N+1$, $j=1,\cdots,N+1$ and $N$ denotes the number of plates. This representation is helpful in understanding the propagation of multiple paths possible in different regions of $z-z'$ space, as can be seen further in Green's function of two and three plates. When concerned with one plate Green's function, the physical interpretation of reflection and transmission coefficients $r_1,t_1$ can be explained by considering a point of source and path of propagation. When the point of observation, $z<z'$ the point of source, the Green's function can be regrouped as \begin{subequations}
\begin{align}
       a_1<z<z', \hspace{4mm} g^{1,H} &= \left(\hbox{e}^{-\kappa (a_1-z)}+r_1^H \hbox{e}^{-\kappa (z-a_1)} \right) \frac{\hbox{e}^{-\kappa(z'-a_1)}}{2 \kappa}, \\
       z<z'<a_1,  \hspace{4mm} g^{1,H} &= \left(t_1^H \hbox{e}^{-\kappa (a_1-z)}\right) \frac{\hbox{e}^{-\kappa(z'-a_1)}}{2 \kappa}.
       \end{align}
\end{subequations}  
The terms outside the parenthesis refer to the source and terms in parenthesis give reflection and transmission amplitudes of the propagator by $R_1^{<,H} \equiv r_1^H$ and $T_1^{<,H} \equiv t_1^H$ (Superscript $<$ denoting $z<z'$ and subscript $1$ in $R,T$ indicating one plate). Similarly, when $z>z'$ the Green's function across the planar interface is 
\begin{subequations}
\begin{align}
       a_1>z>z', \hspace{4mm} g^{1,H} &= \left(\hbox{e}^{-\kappa (z-a_1)}+r_1^H \hbox{e}^{-\kappa (a_1-z)} \right) \frac{\hbox{e}^{-\kappa(a_1-z')}}{2 \kappa}, \\
       z>z'>a_1,  \hspace{4mm} g^{1,H} &= \left(t_1^H \hbox{e}^{-\kappa (z-a_1)}\right) \frac{\hbox{e}^{-\kappa(a_1-z')}}{2 \kappa}.
       \end{align}
\end{subequations}
These give the reflection and transmission amplitudes $R_1^{>,H} \equiv r_1^H$ and $T_1^{>,H} \equiv t_1^H$. These reflection coefficients can also be obtained from matrix $B^H$, where $R_1^{<,H}=B^H_{12}$ and $R_1^{>,H}=B^H_{21}$. And transmission coefficients can be obtained from $B^H_{11}$ and $B^H_{22}$. Transmission amplitudes $T^{<,H}=T^{>,H}$ for all configurations, but $R_1^{<,H}=R_1^{>,H}$ only in the case of one plate, which differ for the case of two plate and three plate reflection amplitudes $R_{12}^{H}, R_{123}^{H}$, as can be seen further. 

Similarly, the solution for $g^{2,H}$ of two plates with the subscript label in Fig.(\ref{1b}) represents the different regions in $z-z'$ space is:
  \begin{subequations}
  \begin{align}
  g^{2,H}_{ \rput(0.1,0){\scriptstyle{11}}
  \pscircle[linewidth=0.2pt](0.1,0.01){0.15} }
  (z,z^\prime) &= \frac{t_1^H \ \hbox{e}^{-\kappa a} \ t_2^H}{2 \kappa \Delta_{12}^H} \hbox{e}^{-\kappa(a_1-z)} \hbox{e}^{-\kappa(z'-a_2)},
  && z<a_1<a_2<z^\prime, \\
  g^{2,H}_{ \rput(0.1,0){\scriptstyle{12}}
  \pscircle[linewidth=0.2pt](0.1,0.01){0.15} }
  (z,z^\prime) &= \frac{t_2^H}{2 \kappa \Delta_{12}^H} \hbox{e}^{-\kappa(a_2-z)} \hbox{e}^{-\kappa(z'-a_2)}+\frac{r_1^H \ \hbox{e}^{-\kappa a} \ t_2^H}{2 \kappa \Delta_{12}^H} \hbox{e}^{-\kappa(z-a_1)} \hbox{e}^{-\kappa(z'-a_2)},
  && a_1<z<a_2<z^\prime, \\
  g^{2,H}_{ \rput(0.1,0){\scriptstyle{13}}
  \pscircle[linewidth=0.2pt](0.1,0.01){0.15} }
  (z,z^\prime) &= \frac{1}{2 \kappa} \hbox{e}^{-\kappa|z-z'|}+\frac{r_2^H}{2 \kappa} \hbox{e}^{-\kappa(z-a_2)} \hbox{e}^{-\kappa(z'-a_2)}+\frac{t_2^H\ \hbox{e}^{-\kappa a} \ r_1^H \ \hbox{e}^{-\kappa a} \ t_2^H}{2 \kappa \Delta_{12}^H} \hbox{e}^{-\kappa(z-a_2)} \hbox{e}^{-\kappa(z'-a_2)},
  && a_1<a_2<z,z^\prime
  \end{align}
  \end{subequations}
  and
  \begin{subequations}
  \begin{align}
  g^{2,H}_{ \rput(0.1,0){\scriptstyle{21}}
  \pscircle[linewidth=0.2pt](0.1,0.01){0.15} }
  (z,z^\prime) &= \frac{t_1^H}{2 \kappa \Delta_{12}^H} \hbox{e}^{-\kappa(a_1-z)} \hbox{e}^{-\kappa(z'-a_1)}+\frac{t_1^H \ \hbox{e}^{-\kappa a} \ r_2^H }{2 \kappa \Delta_{12}^H} \hbox{e}^{-\kappa(a_1-z)} \hbox{e}^{-\kappa(a_2-z')}, 
  && z<a_1<z^\prime<a_2, \\
  g^{2,H}_{ \rput(0.1,0){\scriptstyle{22}}
  \pscircle[linewidth=0.2pt](0.1,0.01){0.15} }
  (z,z^\prime) &= \frac{1}{2 \kappa} \hbox{e}^{-\kappa|z-z'|}+\frac{r_1^H}{2 \kappa \Delta_{12}^H} \hbox{e}^{-\kappa(z-a_1)} \hbox{e}^{-\kappa(z'-a_1)} +\frac{r_1^H \ \hbox{e}^{-\kappa a} \ r_2^H}{2 \kappa \Delta_{12}^H} \hbox{e}^{-\kappa(z-a_1)} \hbox{e}^{-\kappa(a_2-z')}
  \nonumber \\ & \hspace{4mm}
  +\frac{r_2^H}{2 \kappa \Delta_{12}^H} \hbox{e}^{-\kappa(a_2-z)} \hbox{e}^{-\kappa(a_2-z')}+\frac{r_1^H \ \hbox{e}^{-\kappa a} \ r_2^H}{2 \kappa \Delta_{12}^H} \hbox{e}^{-\kappa(a_2-z)} \hbox{e}^{-\kappa(z'-a_1)},
  && a_1<z^\prime,z<a_2, \\
  g^{2,H}_{ \rput(0.1,0){\scriptstyle{23}}
  \pscircle[linewidth=0.2pt](0.1,0.01){0.15} }
  (z,z^\prime) &= \frac{t_2^H}{2 \kappa \Delta_{12}^H} \hbox{e}^{-\kappa(z-a_2)} \hbox{e}^{-\kappa(a_2-z')}+\frac{r_1^H \ \hbox{e}^{-\kappa a} \ t_2^H}{2 \kappa \Delta_{12}^H} \hbox{e}^{-\kappa(z-a_2)} \hbox{e}^{-\kappa(z'-a_1)},
  && a_1<z^\prime<a_2<z
  \end{align}
  \end{subequations}
  and
  \begin{subequations}
  \begin{align}
  g^{2,H}_{ \rput(0.1,0){\scriptstyle{31}}
  \pscircle[linewidth=0.2pt](0.1,0.01){0.15} }
  (z,z^\prime) &= \frac{1}{2 \kappa} \hbox{e}^{-\kappa|z-z'|}+\frac{r_1^H}{2 \kappa} \hbox{e}^{-\kappa(a_1-z)} \hbox{e}^{-\kappa(a_1-z')}+\frac{t_1^H\ \hbox{e}^{-\kappa a} \ r_2^H \ \hbox{e}^{-\kappa a} \ t_1^H}{2 \kappa \Delta_{12}^H} \hbox{e}^{-\kappa(a_1-z)} \hbox{e}^{-\kappa(a_1-z')},
  && z,z^\prime< a_1<a_2, \\
  g^{2,H}_{ \rput(0.1,0){\scriptstyle{32}}
  \pscircle[linewidth=0.2pt](0.1,0.01){0.15} }
  (z,z^\prime) &= \frac{t_1^H}{2 \kappa \Delta_{12}^H} \hbox{e}^{-\kappa(z-a_1)} \hbox{e}^{-\kappa(a_1-z')}+\frac{r_2^H\ \hbox{e}^{-\kappa a} \ t_1^H}{2 \kappa \Delta_{12}^H} \hbox{e}^{-\kappa(a_2-z)} \hbox{e}^{-\kappa(a_1-z')}, && z^\prime<a_1<z<a_2,\\
  g^{2,H}_{ \rput(0.1,0){\scriptstyle{33}}
  \pscircle[linewidth=0.2pt](0.1,0.01){0.15} }
  (z,z^\prime) &= \frac{t_2^H \ \hbox{e}^{-\kappa a} \ t_1^H}{2 \kappa \Delta_{12}^H} \hbox{e}^{-\kappa(z-a_2)} \hbox{e}^{-\kappa(a_1-z')},
  && z^\prime<a_1<a_2<z.
  \end{align}
  \end{subequations} 
The reflection and transmission coefficients at each plate are defined in Eq. (\ref{coeff}) for $i=1,2$. Multiple scattering parameter here is \begin{equation}
    \Delta_{12}^H=1-r_1^H \hbox{e}^{-\kappa a} r_2^H \hbox{e}^{-\kappa a},
    \label{MS1}
 \end{equation} with $a=a_2-a_1$ is the distance between the plates. Similar to matrices in Eq. (\ref{mat}), this Green's function $g^{2,H}$ can be represented by constructing matrices $A$, $B^H$ and $C$ such as \begin{multline}
     A = \left[\begin{array}{c c c}
  \hbox{e}^{-\kappa(z'-a_2)} &  \left[\begin{array}{c c}
   \hbox{e}^{-\kappa(z'-a_1)} &  \hbox{e}^{-\kappa(a_2-z')} \end{array}\right] &  \hbox{e}^{-\kappa(a_1-z')} \end{array}\right], \\ B^H = \left[\begin{array}{c c c} \frac{t_1^H \hbox{e}^{-\kappa a} t_2^H}{\Delta_{12}^H} & \left[\begin{array}{c c}
   \frac{t_2^H}{\Delta_{12}^H} \hspace{1.5mm} & \hspace{1.5mm} \frac{r_1^H \hbox{e}^{-\kappa a} t_2^H}{\Delta_{12}^H} \end{array}\right] &  r_2^H +\frac{t_2^H\hbox{e}^{-\kappa a}r_1^H\hbox{e}^{-\kappa a}t_2^H}{\Delta_{12}^H} \vspace{3mm}  \\ \left[\begin{array}{c c}
  \frac{t_1^H}{\Delta_{12}^H} \vspace{3mm} \\ \vspace{3mm}\frac{t_1^H \hbox{e}^{-\kappa a} r_2^H}{\Delta_{12}^H} \end{array}\right] & \left[\begin{array}{c c} \frac{r_1^H \hbox{e}^{-\kappa a} r_2^H}{\Delta_{12}^H} & \frac{r_1^H}{\Delta_{12}^H} \vspace{3mm} \\ \vspace{3mm} \frac{r_2^H}{\Delta_{12}^H} & \frac{r_2^H \hbox{e}^{-\kappa a} r_1^H}{\Delta_{12}^H}
  \end{array}\right] & \left[\begin{array}{c c}
   \frac{r_1^H \hbox{e}^{-\kappa a} t_2^H}{\Delta_{12}^H} \vspace{3mm} \\ \vspace{3mm} \frac{t_2^H}{\Delta_{12}^H} \end{array}\right] \vspace{3mm} \\ \vspace{3mm} r_1^H +\frac{t_1^H\hbox{e}^{-\kappa a}r_2^H\hbox{e}^{-\kappa a}t_1^H}{\Delta_{12}^H} &\left[\begin{array}{c c}
  \frac{r_2^H\hbox{e}^{-\kappa a}t_1^H}{\Delta_{12}^H} &  \frac{t_1^H}{\Delta_{12}^H} \end{array}\right] &  \frac{t_2^H \hbox{e}^{-\kappa a} t_1^H}{\Delta_{12}^H}
  \end{array}\right]\\\hbox{and}\, C = \left[\begin{array}{c c c}
  \hbox{e}^{-\kappa(a_1-z)} \\ \left[\begin{array}{c c}
  \hbox{e}^{-\kappa(a_2-z)} \\ \hbox{e}^{-\kappa(z-a_1)} \end{array}\right] \\  \hbox{e}^{-\kappa(z-a_2)} \end{array}\right].
  \end{multline}
  Different regions in Green's function $ g^{2,H}_{ \rput(0.1,0){\scriptstyle{ij}}
 \pscircle[linewidth=0.2pt](0.1,0.01){0.15} }$ can 
be obtained here from Eq. (\ref{matrix}) for $N=2$. The reflection coefficients of two plates configuration can be obtained here from  $R_{12}^{<,H}=B^H_{13}$ and $R_{12}^{>,H}=B^H_{31}$. And transmission coefficients of two plates configuration are $T_{12}^{<,H}=T_{12}^{>,H}=B^H_{11}=B^H_{33}$. The remaining terms describe the multiple possible ways for the path of propagation depending on $z$ and $z'$ with an exponential dependence on the length of propagation. The reflection coefficients are not the same when $z<z'$ and $z>z'$, but the path of propagation defined in the terms $B^H_{11}$, $B^H_{12}$ and $B^H_{21}$ are same as $B^H_{33}$, $B^H_{23}$ and $B^H_{32}$, respectively where the matrix is symmetric around the anti-diagonal terms ${B^H_{ij}}$ for $i+j-2=N$.  \\
  The solution for $g^{3,H}$ of three plates with the subscript label in Fig. (\ref{Fig2}) representing the different regions in $z-z'$ space is:
  \begin{subequations}
       \begin{align}
         g^{3,H}_{ \rput(0.1,0){\scriptstyle{11}}
  \pscircle[linewidth=0.2pt](0.1,0.01){0.15} }
  (z,z^\prime) &= \frac{t_{1}^H \ \hbox{e}^{-\kappa a} \ t_{2}^H \ \hbox{e}^{-\kappa b} \ t_{3}^H}{2 \kappa \Delta_{123}^H} \hbox{e}^{-\kappa(a_1-z)} \hbox{e}^{-\kappa(z'-a_3)},  \hspace{2mm}
 z<a_1<a_2<a_3<z', \\
  g^{3,H}_{ \rput(0.1,0){\scriptstyle{12}}
  \pscircle[linewidth=0.2pt](0.1,0.01){0.15} }
  (z,z^\prime) &= \frac{t_{2}^H \ \hbox{e}^{-\kappa b} \ t_{3}^H}{2 \kappa \Delta_{123}^H} \hbox{e}^{-\kappa(a_2-z)} \hbox{e}^{-\kappa(z'-a_3)}+\frac{r_{1}^H \ \hbox{e}^{-\kappa a} \ t_{2}^H \ \hbox{e}^{-\kappa b} \ t_{3}^H}{2 \kappa \Delta_{123}^H} \hbox{e}^{-\kappa(z-a_1)} \hbox{e}^{-\kappa(z'-a_3)}, \hspace{2mm}
  a_1<z<a_2<a_3<z', \\
   g^{3,H}_{ \rput(0.1,0){\scriptstyle{13}}
  \pscircle[linewidth=0.2pt](0.1,0.01){0.15} }
  (z,z^\prime) &= \frac{t_{3}^H \ \Delta_{12}^H}{2 \kappa \Delta_{123}^H} \hbox{e}^{-\kappa(a_3-z)} \hbox{e}^{-\kappa(z'-a_3)}+\frac{r_{2}^H\ \hbox{e}^{-\kappa b} \ t_{3}^H \ \Delta_{12}^H}{2 \kappa \Delta_{123}^H} \hbox{e}^{-\kappa(z-a_2)} \hbox{e}^{-\kappa(z'-a_3)} \nonumber \\ & \hspace{4mm} +\frac{t_{2}^H\ \hbox{e}^{-\kappa a} \ r_{1}^H \ \hbox{e}^{-\kappa a} \ t_{2}^H \ \hbox{e}^{-\kappa b} \ t_{3}^H }{2 \kappa \Delta_{123}^H} \hbox{e}^{-\kappa(z-a_2)} \hbox{e}^{-\kappa(z'-a_3)}, \hspace{2mm}
  a_1<a_2<z<a_3<z', \\
   g^{3,H}_{ \rput(0.1,0){\scriptstyle{14}}
  \pscircle[linewidth=0.2pt](0.1,0.01){0.15} }
  (z,z^\prime) &= \frac{1}{2 \kappa} \hbox{e}^{-\kappa|z-z'|}+\frac{r_{3}^H}{2 \kappa} \hbox{e}^{-\kappa(z-a_3)} \hbox{e}^{-\kappa(z'-a_3)}+\frac{t_{3}^H \ \hbox{e}^{-\kappa b} \ r_{2}^H \ \hbox{e}^{-\kappa b} \ t_{3}^H \ \Delta_{12}^H}{2 \kappa \Delta_{123}^H} \hbox{e}^{-\kappa(z-a_3)} \hbox{e}^{-\kappa(z'-a_3)} \nonumber \\ & \hspace{4mm} +\frac{t_{3}^H \ \hbox{e}^{-\kappa b} \ t_{2}^H \ \hbox{e}^{-\kappa a} \ r_{1}^H \ \hbox{e}^{-\kappa a} \ t_{2}^H \ \hbox{e}^{-\kappa b} \ t_{3}^H}{2 \kappa \Delta_{123}^H} \hbox{e}^{-\kappa(z-a_3)} \hbox{e}^{-\kappa(z'-a_3)}, \hspace{2mm}
 a_1<a_2<a_3<z,z'.
      \end{align}
  \end{subequations}
 and
 \begin{subequations}
     \begin{align}
        g^{3,H}_{ \rput(0.1,0){\scriptstyle{21}}
 \pscircle[linewidth=0.2pt](0.1,0.01){0.15} }
 (z,z^\prime) &= \frac{t_{1}^H \ \hbox{e}^{-\kappa a} \ t_{2}^H}{2 \kappa \Delta^H} \hbox{e}^{-\kappa(a_1-z)} \hbox{e}^{-\kappa(z'-a_2)}+\frac{t_{1}^H \ \hbox{e}^{-\kappa a} \ t_{2}^H \ \hbox{e}^{-\kappa b} \ r_{3}^H}{2 \kappa \Delta_{123}^H} \hbox{e}^{-\kappa(a_1-z)} \hbox{e}^{-\kappa(a_3-z')},  \hspace{2mm}
 z<a_1<a_2<z'<a_3, \\
 g^{3,H}_{ \rput(0.1,0){\scriptstyle{22}}
 \pscircle[linewidth=0.2pt](0.1,0.01){0.15} }
 (z,z^\prime) &= \frac{t_{2}^H}{2 \kappa \Delta_{123}^H} \hbox{e}^{-\kappa(a_2-z)} \hbox{e}^{-\kappa(z'-a_2)}+\frac{t_{2}^H \ \hbox{e}^{-\kappa b} \ r_{3}^H}{2 \kappa \Delta_{123}^H} \hbox{e}^{-\kappa(a_2-z)} \hbox{e}^{-\kappa(a_3-z')} \nonumber \\ & \hspace{4mm} +\frac{r_{1}^H \ \hbox{e}^{-\kappa a} \ t_{2}^H}{2 \kappa \Delta_{123}^H} \hbox{e}^{-\kappa(z-a_1)} \hbox{e}^{-\kappa(z'-a_2)}+\frac{r_{1}^H \ \hbox{e}^{-\kappa a} \ t_{2}^H \ \hbox{e}^{-\kappa b} \ r_{3}^H}{2 \kappa \Delta_{123}^H} \hbox{e}^{-\kappa(z-a_1)} \hbox{e}^{-\kappa(a_3-z')},  \hspace{2mm}
 a_1<z<a_2<z'<a_3, \\
 g^{3,H}_{ \rput(0.1,0){\scriptstyle{23}}
 \pscircle[linewidth=0.2pt](0.1,0.01){0.15} }
 (z,z^\prime) &= \frac{1}{2 \kappa} \hbox{e}^{-\kappa|z-z'|}+\frac{r_{3}^H \ \Delta_{12}^H}{2 \kappa \Delta_{123}^H} \hbox{e}^{-\kappa(a_3-z)} \hbox{e}^{-\kappa(a_3-z')}  \nonumber \\ & \hspace{4mm} +\frac{r_{3}^H \ \hbox{e}^{-\kappa b} \ r_{2}^H \Delta_{12}^H}{2 \kappa \Delta_{123}^H} \hbox{e}^{-\kappa(a_3-z)} \hbox{e}^{-\kappa(z'-a_2)}+\frac{r_{3}^H \ \hbox{e}^{-\kappa b} \ t_{2}^H \ \hbox{e}^{-\kappa a} \ r_{1}^H  \ \hbox{e}^{-\kappa a} \ t_{2}^H}{2 \kappa \Delta_{123}^H} \hbox{e}^{-\kappa(a_3-z)} \hbox{e}^{-\kappa(z'-a_2)}  \nonumber \\ & \hspace{4mm} +\frac{t_{2}^H \ \hbox{e}^{-\kappa a} \ r_{1}^H \ \hbox{e}^{-\kappa a} \ t_{2}^H  \ \hbox{e}^{-\kappa b} \ r_{3}^H}{2 \kappa \Delta_{123}^H} \hbox{e}^{-\kappa(z-a_2)} \hbox{e}^{-\kappa(a_3-z')}+\frac{r_{2}^H \ \hbox{e}^{-\kappa b} \ r_{3}^H  \ \Delta_{12}^H}{2 \kappa \Delta_{123}^H} \hbox{e}^{-\kappa(z-a_2)} \hbox{e}^{-\kappa(a_3-z')} \nonumber \\ & \hspace{4mm} +\frac{r_{2}^H \ \Delta_{12}^H}{2 \kappa \Delta_{123}^H} \hbox{e}^{-\kappa(z-a_2)} \hbox{e}^{-\kappa(z'-a_2)}+\frac{t_{2}^H \ \hbox{e}^{-\kappa a} \ r_{1}^H \ \hbox{e}^{-\kappa a} \ t_{2}^H}{2 \kappa \Delta_{123}^H} \hbox{e}^{-\kappa(z-a_2)} \hbox{e}^{-\kappa(z'-a_2)},  \hspace{2mm}
 a_1<a_2<z,z'<a_3, \\
 g^{3,H}_{ \rput(0.1,0){\scriptstyle{24}}
 \pscircle[linewidth=0.2pt](0.1,0.01){0.15} }
 (z,z^\prime) &= \frac{t_{3}^H \ \Delta_{12}^H}{2 \kappa \Delta_{123}^H} \hbox{e}^{-\kappa(z-a_3)} \hbox{e}^{-\kappa(a_3-z')}+\frac{t_{3}^H\ \hbox{e}^{-\kappa b} \ r_{2}^H \ \Delta_{12}^H}{2 \kappa \Delta_{123}^H} \hbox{e}^{-\kappa(z-a_3)} \hbox{e}^{-\kappa(z'-a_2)} \nonumber \\ & \hspace{4mm} +\frac{t_{3}^H\ \hbox{e}^{-\kappa b} \ t_{2}^H \ \hbox{e}^{-\kappa a} \ r_{1}^H \ \hbox{e}^{-\kappa a} \ t_{2}^H }{2 \kappa \Delta_{123}^H} \hbox{e}^{-\kappa(z-a_3)} \hbox{e}^{-\kappa(z'-a_2)},  \hspace{2mm}
 a_1<a_2<z'<a_3<z,
     \end{align}
 \end{subequations}
 and  
 \begin{subequations}
     \begin{align}
         g^{3,H}_{ \rput(0.1,0){\scriptstyle{31}}
 \pscircle[linewidth=0.2pt](0.1,0.01){0.15} }
 (z,z^\prime) &= \frac{t_{1}^H \ \Delta_{23}^H}{2 \kappa \Delta_{123}^H} \hbox{e}^{-\kappa(a_1-z)} \hbox{e}^{-\kappa(z'-a_1)}+\frac{t_{1}^H\ \hbox{e}^{-\kappa a} \ r_{2}^H \ \Delta_{23}^H}{2 \kappa \Delta_{123}^H} \hbox{e}^{-\kappa(a_1-z)} \hbox{e}^{-\kappa(a_2-z')} \nonumber \\ & \hspace{4mm} +\frac{t_{1}^H\ \hbox{e}^{-\kappa a} \ t_{2}^H \ \hbox{e}^{-\kappa b} \ r_{3}^H \ \hbox{e}^{-\kappa b} \ t_{2}^H }{2 \kappa \Delta_{123}^H} \hbox{e}^{-\kappa(a_1-z)} \hbox{e}^{-\kappa(a_2-z')}, \hspace{2mm}
 z<a_1<z'<a_2<a_3, \\ 
  g^{3,H}_{ \rput(0.1,0){\scriptstyle{32}}
 \pscircle[linewidth=0.2pt](0.1,0.01){0.15} }
 (z,z^\prime) &= \frac{1}{2 \kappa} \hbox{e}^{-\kappa|z-z'|}+\frac{r_{1}^H \ \Delta_{23}^H}{2 \kappa \Delta_{123}^H} \hbox{e}^{-\kappa(z-a_1)} \hbox{e}^{-\kappa(z'-a_1)} \nonumber \\ & \hspace{4mm} +\frac{r_{1}^H \ \hbox{e}^{-\kappa a} \ r_{2}^H \Delta_{23}^H}{2 \kappa \Delta_{123}^H} \hbox{e}^{-\kappa(z-a_1)} \hbox{e}^{-\kappa(a_2-z')}+\frac{r_{1}^H \ \hbox{e}^{-\kappa a} \ t_{2}^H \ \hbox{e}^{-\kappa b} \ r_{3}^H  \ \hbox{e}^{-\kappa b} \ t_{2}^H}{2 \kappa \Delta_{123}^H} \hbox{e}^{-\kappa(z-a_1)} \hbox{e}^{-\kappa(a_2-z')} \nonumber \\ & \hspace{4mm} +\frac{t_{2}^H \ \hbox{e}^{-\kappa b} \ r_{3}^H \ \hbox{e}^{-\kappa b} \ t_{2}^H  \ \hbox{e}^{-\kappa a} \ r_{1}^H}{2 \kappa \Delta_{123}^H} \hbox{e}^{-\kappa(a_2-z)} \hbox{e}^{-\kappa(z'-a_1)}+\frac{r_{2}^H \ \hbox{e}^{-\kappa a} \ r_{1}^H  \ \Delta_{23}^H}{2 \kappa \Delta_{123}^H} \hbox{e}^{-\kappa(a_2-z)} \hbox{e}^{-\kappa(z'-a_1)} \nonumber \\ & \hspace{4mm} +\frac{r_{2}^H \ \Delta_{23}^H}{2 \kappa \Delta_{123}^H} \hbox{e}^{-\kappa(a_2-z)} \hbox{e}^{-\kappa(a_2-z')}+\frac{t_{2}^H \ \hbox{e}^{-\kappa b} \ r_{3}^H \ \hbox{e}^{-\kappa b} \ t_{2}^H}{2 \kappa \Delta_{123}^H} \hbox{e}^{-\kappa(a_2-z)} \hbox{e}^{-\kappa(a_2-z')}, \hspace{2mm}
 a_1<z,z'<a_2<a_3, \\ 
 g^{3,H}_{ \rput(0.1,0){\scriptstyle{33}}
 \pscircle[linewidth=0.2pt](0.1,0.01){0.15} }
 (z,z^\prime) &= \frac{t_{2}^H}{2 \kappa \Delta_{123}^H} \hbox{e}^{-\kappa(z-a_2)} \hbox{e}^{-\kappa(a_2-z')}+\frac{t_{2}^H \ \hbox{e}^{-\kappa a} \ r_{1}^H}{2 \kappa \Delta_{123}^H} \hbox{e}^{-\kappa(z-a_2)} \hbox{e}^{-\kappa(z'-a_1)} \nonumber \\ & \hspace{4mm} +\frac{r_{3}^H \ \hbox{e}^{-\kappa b} \ t_{2}^H}{2 \kappa \Delta_{123}^H} \hbox{e}^{-\kappa(a_3-z)} \hbox{e}^{-\kappa(a_2-z')}+\frac{r_{3}^H \ \hbox{e}^{-\kappa b} \ t_{2}^H \ \hbox{e}^{-\kappa a} \ r_{1}^H}{2 \kappa \Delta_{123}^H} \hbox{e}^{-\kappa(a_3-z)} \hbox{e}^{-\kappa(z'-a_1)}, \hspace{2mm}
 a_1<z'<a_2<z<a_3, \\
 g^{3,H}_{ \rput(0.1,0){\scriptstyle{34}}
 \pscircle[linewidth=0.2pt](0.1,0.01){0.15} }
 (z,z^\prime) &= \frac{t_{3}^H \ \hbox{e}^{-\kappa b} \ t_{2}^H}{2 \kappa \Delta_{123}^H} \hbox{e}^{-\kappa(z-a_3)} \hbox{e}^{-\kappa(a_2-z')}+\frac{t_{3}^H \ \hbox{e}^{-\kappa b} \ t_{2}^H \ \hbox{e}^{-\kappa a} \ r_{1}^H}{2 \kappa \Delta_{123}^H} \hbox{e}^{-\kappa(z-a_3)} \hbox{e}^{-\kappa(z'-a_1)}, \hspace{2mm}
 a_1<z'<a_2<a_3<z,
 \end{align}
 \end{subequations}
  and \begin{subequations}
     \begin{align}
  g^{3,H}_{ \rput(0.1,0){\scriptstyle{41}}
 \pscircle[linewidth=0.2pt](0.1,0.01){0.15} }
 (z,z^\prime) &= \frac{1}{2 \kappa} \hbox{e}^{-\kappa|z-z'|}+\frac{r_{1}^H}{2 \kappa} \hbox{e}^{-\kappa(a_1-z)} \hbox{e}^{-\kappa(a_1-z')}+\frac{t_{1}^H \ \hbox{e}^{-\kappa a} \ r_{2}^H \ \hbox{e}^{-\kappa a} \ t_{1}^H \ \Delta_{23}^H}{2 \kappa \Delta_{123}^H} \hbox{e}^{-\kappa(a_1-z)} \hbox{e}^{-\kappa(a_1-z')} \nonumber \\ & \hspace{4mm} +\frac{t_{1}^H \ \hbox{e}^{-\kappa a} \ t_{2}^H \ \hbox{e}^{-\kappa b} \ r_{3}^H \ \hbox{e}^{-\kappa b} \ t_{2}^H \ \hbox{e}^{-\kappa a} \ t_{1}^H}{2 \kappa \Delta_{123}^H} \hbox{e}^{-\kappa(a_1-z)} \hbox{e}^{-\kappa(a_1-z')}, \hspace{2mm} z,z^\prime< a_1<a_2<a_3, \\
 g^{3,H}_{ \rput(0.1,0){\scriptstyle{42}}
 \pscircle[linewidth=0.2pt](0.1,0.01){0.15} }
 (z,z^\prime) &=\frac{t_{1}^H \ \Delta_{23}^H}{2 \kappa \Delta_{123}^H} \hbox{e}^{-\kappa(z-a_1)} \hbox{e}^{-\kappa(a_1-z')}+\frac{r_{2}^H\ \hbox{e}^{-\kappa a} \ t_{1}^H \ \Delta_{23}^H}{2 \kappa \Delta_{123}^H} \hbox{e}^{-\kappa(a_2-z)} \hbox{e}^{-\kappa(a_1-z')} \nonumber \\ & \hspace{4mm} +\frac{t_{2}^H\ \hbox{e}^{-\kappa b} \ r_{3}^H \ \hbox{e}^{-\kappa b} \ t_{2}^H \ \hbox{e}^{-\kappa a} \ t_{1}^H }{2 \kappa \Delta_{123}^H} \hbox{e}^{-\kappa(a_2-z)} \hbox{e}^{-\kappa(a_1-z')}, \hspace{2mm} z^\prime<a_1<z<a_2<a_3, \\
  g^{3,H}_{ \rput(0.1,0){\scriptstyle{43}}
 \pscircle[linewidth=0.2pt](0.1,0.01){0.15} }
 (z,z^\prime) &= \frac{t_{2}^H \ \hbox{e}^{-\kappa a} \ t_{1}^H}{2 \kappa \Delta_{123}^H} \hbox{e}^{-\kappa(z-a_2)} \hbox{e}^{-\kappa(a_1-z')}+\frac{r_{3}^H \ \hbox{e}^{-\kappa b} \ t_{2}^H \ \hbox{e}^{-\kappa a} \ t_{1}^H}{2 \kappa \Delta_{123}^H} \hbox{e}^{-\kappa(a_3-z)} \hbox{e}^{-\kappa(a_1-z')}, \hspace{2mm} z^\prime<a_1<a_2<z<a_3, \\
 g^{3,H}_{ \rput(0.1,0){\scriptstyle{44}}
 \pscircle[linewidth=0.2pt](0.1,0.01){0.15} }
 (z,z^\prime) &= \frac{t_{3}^H \ \hbox{e}^{-\kappa b} \ t_{2}^H \ \hbox{e}^{-\kappa a} \ t_{1}^H}{2 \kappa \Delta_{123}^H} \hbox{e}^{-\kappa(z-a_3)} \hbox{e}^{-\kappa(a_1-z')}, \hspace{2mm}
 z^\prime<a_1<a_2<a_3<z.
 \end{align}
 \end{subequations} 
  Matrices $A$, $B^H$ and $C$ can be defined for this cumbersome expression of Green's function as \begin{multline}
     A = \left[\begin{array}{c c c c}
\hbox{e}^{-\kappa(z'-a_3)} &  \left[\begin{array}{c c} \hbox{e}^{-\kappa(z'-a_2)} &  \hbox{e}^{-\kappa(a_3-z')} \end{array}\right] &  \left[\begin{array}{c c}
  \hbox{e}^{-\kappa(z'-a_1)} &  \hbox{e}^{-\kappa(a_2-z')} \end{array}\right] &  \hbox{e}^{-\kappa(a_1-z')} \end{array}\right], \\ B^H = \frac{1}{\Delta_{123}^H} \left[\begin{array}{cccc}
 \substack{t_{1}^H \hbox{e}^{-\kappa a} t_{2}^H \hbox{e}^{-\kappa b} t_{3}^H }  &  B'_{12} & B'_{13} & \substack{ r_{3}^H \Delta_{123}^H + t_{3}^H \hbox{e}^{-\kappa b} r_{2}^H \hbox{e}^{-\kappa b} t_{3}^H \Delta_{12}^H \\ + t_{3}^H \hbox{e}^{-\kappa b} t_{2}^H \hbox{e}^{-\kappa a} r_{1}^H \hbox{e}^{-\kappa a} t_{2}^H \hbox{e}^{-\kappa b} t_{3}^H}  \\ 
 B'_{21}   &  B'_{22} & B'_{23} & B'_{24} \\
 B'_{31}   &  B'_{32} & B'_{33} & B'_{34}  \\  \substack{
 r_1^H \Delta_{123}^H + t_1^H \hbox{e}^{-\kappa a} r_2^H \hbox{e}^{-\kappa a} t_1^H \Delta_{23}^H  \\ + t_1^H \hbox{e}^{-\kappa a} t_2^H \hbox{e}^{-\kappa b} r_3^H \hbox{e}^{-\kappa b} t_2^H \hbox{e}^{-\kappa a} t_1^H}  &  B'_{42} & B'_{43} & \substack{ t_{3}^H \hbox{e}^{-\kappa b} t_{2}^H \hbox{e}^{-\kappa a} t_{1}^H }
 \end{array}\right] \\\hbox{and}\, C = \left[\begin{array}{c c c c}
 \hbox{e}^{-\kappa(a_1-z)} \\ \left[\begin{array}{c c}
 \hbox{e}^{-\kappa(a_2-z)} \\ \hbox{e}^{-\kappa(z-a_1)} \end{array}\right]  \vspace{3mm} \\  \vspace{3mm} \left[\begin{array}{c c}
 \hbox{e}^{-\kappa(a_3-z)}  \\  
  \hbox{e}^{-\kappa(z-a_2)} \end{array}\right] \\ \hbox{e}^{-\kappa(z-a_3)} \end{array}\right]
\end{multline}
 with
 \begin{equation}
     B'_{12}=\left[\begin{array}{c c} \substack{t_{2}^H  \hbox{e}^{-\kappa b} t_{3}^H} \hspace{3mm} & \hspace{3mm} \substack{ r_{1}^H \hbox{e}^{-\kappa a} t_{2}^H \hbox{e}^{-\kappa b} \ t_{3}^H } \end{array}\right],\,B'_{34}=\left[\begin{array}{c c} \substack{t_{3}^H \hbox{e}^{-\kappa b} t_{2}^H \hbox{e}^{-\kappa a} r_{1}^H} \\ \substack{t_{3}^H  \hbox{e}^{-\kappa b} t_{2}^H} \end{array}\right],
 \end{equation}
 \begin{equation}
    B'_{13}=\left[\begin{array}{c c}\substack{t_{3}^H \Delta_{12}^H} \hspace{3mm} &  \hspace{3mm} \substack{r_{2}^H \hbox{e}^{-\kappa b}  t_{3}^H \Delta_{12}^H + t_{2}^H\ \hbox{e}^{-\kappa a} r_{1}^H \hbox{e}^{-\kappa a} t_{2}^H \hbox{e}^{-\kappa b} t_{3}^H } \end{array}\right],\,B'_{24}=\left[\begin{array}{c c} \substack{ t_{3}^H \hbox{e}^{-\kappa b} r_{2}^H \Delta_{12}^H + t_{3}^H \hbox{e}^{-\kappa b}  t_{2}^H  \hbox{e}^{-\kappa a}  r_{1}^H \hbox{e}^{-\kappa a} t_{2}^H } \\ \substack{t_{3}^H \Delta_{12}^H } \end{array}\right],
 \end{equation}
 \begin{equation}
   B'_{21}=\left[\begin{array}{c c}
\substack{t_{1}^H \hbox{e}^{-\kappa a} t_{2}^H} \\ \substack{t_{1}^H \hbox{e}^{-\kappa a} t_{2}^H \hbox{e}^{-\kappa b} r_{3}^H} \end{array}\right],\,B'_{43}=\left[\begin{array}{c c} \substack{r_{3}^H \hbox{e}^{-\kappa b} t_{2}^H \hbox{e}^{-\kappa a} t_{1}^H} \hspace{3mm} & \hspace{3mm} \substack{ t_{2}^H \hbox{e}^{-\kappa a} t_{1}^H}\end{array}\right],
 \end{equation}
 \begin{equation}
     B'_{22}=\left[\begin{array}{c c} \substack{ t_{2}^H} & \substack{r_{1}^H \hbox{e}^{-\kappa a} t_{2}^H} \\ \substack{t_{2}^H \hbox{e}^{-\kappa b} r_{3}^H} & \substack{ r_{1}^H \hbox{e}^{-\kappa a} t_{2}^H \hbox{e}^{-\kappa b} r_{3}^H }
\end{array}\right],\,B'_{33}=\left[\begin{array}{c c} \substack{r_{3}^H \hbox{e}^{-\kappa b} t_{2}^H \hbox{e}^{-\kappa a} r_{1}^H} & \substack{t_{2}^H \hbox{e}^{-\kappa a} r_{1}^H} \\ \substack{r_{3}^H \hbox{e}^{-\kappa b} t_{2}^H} & \substack{t_{2}^H}
 \end{array}\right],
 \end{equation}
 \begin{equation}
     B'_{31}=\left[\begin{array}{c c}
 \substack{t_{1}^H \Delta_{23}^H}\\ \substack{t_{1}^H \hbox{e}^{-\kappa a}  r_{2}^H  \Delta_{23}^H + t_{1}^H \hbox{e}^{-\kappa a}  t_{2}^H  \hbox{e}^{-\kappa b}  r_{3}^H  \hbox{e}^{-\kappa b}  t_{2}^H} \end{array}\right],\,B'_{42}=\left[\begin{array}{c c} \substack{t_{1}^H \Delta_{23}^H} \hspace{3mm} &  \hspace{3mm} \substack{r_{2}^H \hbox{e}^{-\kappa a}  t_{1}^H \Delta_{23}^H +
 t_{2}^H \hbox{e}^{-\kappa b} r_{3}^H \hbox{e}^{-\kappa b} t_{2}^H \hbox{e}^{-\kappa a} t_{1}^H} \end{array}\right],
 \end{equation}
 \begin{equation}
     B'_{23}= \left[\begin{array}{c c} \substack{r_{3}^H \hbox{e}^{-\kappa b} r_{2}^H \Delta_{12}^H + r_{3}^H \hbox{e}^{-\kappa b} t_{2}^H \hbox{e}^{-\kappa a} r_{1}^H \hbox{e}^{-\kappa a} t_{2}^H } \hspace{3mm} & \hspace{3mm} \substack{ r_{2}^H \Delta_{12}^H+ t_{2}^H \hbox{e}^{-\kappa a} r_{1}^H \hbox{e}^{-\kappa a} t_{2}^H } \\ \substack{ r_{3}^H  \Delta_{12}^H } & \substack{ t_{2}^H \hbox{e}^{-\kappa a} r_{1}^H \hbox{e}^{-\kappa a} t_{2}^H \hbox{e}^{-\kappa b} r_{3}^H+r_{2}^H \hbox{e}^{-\kappa b} r_{3}^H  \Delta_{12}^H }
 \end{array}\right]
 \end{equation} and 
 \begin{equation}
     B'_{32}= \left[\begin{array}{c c} \substack{t_{2}^H \hbox{e}^{-\kappa b} r_{3}^H \hbox{e}^{-\kappa b} t_{2}^H \hbox{e}^{-\kappa a} r_{1}^H+r_{2}^H \hbox{e}^{-\kappa a} r_{1}^H \Delta_{23}^H} & \substack{r_{1}^H  \Delta_{23}^H} \\ \substack{r_{2}^H \Delta_{23}^H + t_{2}^H \hbox{e}^{-\kappa b} r_{3}^H \hbox{e}^{-\kappa b} t_{2}^H} & \substack{ r_{1}^H \hbox{e}^{-\kappa a} r_{2}^H \Delta_{23}^H + r_{1}^H \hbox{e}^{-\kappa a} t_{2}^H \hbox{e}^{-\kappa b} r_{3}^H \hbox{e}^{-\kappa b} t_{2}^H}
 \end{array}\right].
 \end{equation}
Different regions in Green's function $ g^{3,H}_{ \rput(0.1,0){\scriptstyle{ij}}
 \pscircle[linewidth=0.2pt](0.1,0.01){0.15} }$ can 
be obtained here from Eq. (\ref{matrix}) for $N=3$. Further, multiple scattering parameters here are \begin{equation}
              \Delta_{123}^H=(\Delta_{12}^H\Delta_{23}^H-r_1^H \hbox{e}^{-\kappa a}  t_2^H \hbox{e}^{-\kappa b} r_3^H \hbox{e}^{-\kappa b} t_2^H \hbox{e}^{-\kappa a})\,\hbox{and}\,\Delta_{23}^H=1-r_2^H \hbox{e}^{-\kappa b} r_3^H \hbox{e}^{-\kappa b}, \label{MS3}
              \end{equation} 
               with $b=a_3-a_2$ is the distance between the plates. The relevance of these two plates multiple scattering parameters $\Delta_{12}^H$, $\Delta_{23}^H$ and three plates multiple scattering parameter $\Delta_{123}^H$ can be seen further in the evaluation of energy (Sec. \rom{5}), where they give all paths of propagation contributing to the energy. The reflection coefficients of three plates configuration can be obtained here from  $R_{123}^{<,H}=B^H_{14}$ and $R_{123}^{>,H}=B^H_{41}$. And transmission coefficients of three plates configuration here are $T_{123}^{<,H}=T_{123}^{>,H}=B^H_{11}=B^H_{44}$.  
               
               Similarly, $g^E$ can be obtained for these configurations by solving differential equation in Eq. (\ref{18}) and using boundary conditions in Eq. (\ref{37}). However, $g^E$ have similar expressions as $g^H$, obtained by replacing superscripts $H \rightarrow E$ and by swapping $\lambda^{\perp}_{e i} \leftrightarrow \lambda^{\perp}_{g i}$ as can be deduced from the boundary conditions.
\section{CASIMIR FORCE FROM STRESS TENSOR METHOD}
Here, we are interested in calculating the force on a plate in two plates and three plates configurations from stress tensor formalism that is based on conservation of electromagnetic momentum\,\cite{schwinger1998classical}. This is calculated using their Green's functions solved in previous section and relating them with their Green's dyadic from Eq. (\ref{30}).  From the Maxwell's equations in Eq. (\ref{3}) one can derive the force density ${\bf f}$ across the surface of a volume $V$, which measures the flux of momentum density of the electromagnetic field when integrated over volume $V$ from implementing the Gauss law. This stress from electromagnetic radiation on the surface of volume $V$ is given as \begin{equation}
{\bf f} = -{\bm\nabla}\cdot {\bf T}. 
\label{for-rad}
\end{equation}
Here $\bf T$ is the stress tensor constituting the force of a plate at $z=a_i$, \begin{equation}
{\bf T} = {\bf 1} \frac{1}{2} ({\bf D} \cdot {\bf E} + {\bf B} \cdot {\bf H})
- ({\bf D} {\bf E} + {\bf B} {\bf H})
\label{stten-def}
\end{equation}
and integration volume $V$ can be represented by a thin film with surfaces $z=a_i-\delta$ and $z=a_i+\delta$ enclosing the plate, by taking the limit $\delta \rightarrow 0$. The force on the surface of plate is given as \begin{equation}
{\bf F}
= -\frac{1}{\tau} \int_{-\infty}^\infty \frac{d\omega}{2\pi} 
\oint_V d{\bf S} \cdot {\bf T}({\bf r}, \omega),
\end{equation} for $\tau=2T\to\infty$. The fluctuations of quantum vacuum dictate that mean value of the field $\langle {\bf E} \rangle =0$ do not contribute and only bilinear description of fields in Eq. (\ref{cor-fields-GD}) contribute to the flux tensor. The Casimir force, a manifestation of quantum vacuum from flux tensor is obtained as \begin{equation}
{\bf F}
= -\frac{1}{2T} \int_{-\infty}^\infty \frac{d\omega}{2\pi}
\oint_V d{\bf S} \cdot \langle {\bf T}({\bf r}, \omega) \rangle.
\label{casfor-def}
\end{equation}
The Casimir pressure or force per unit area on the plate at $z=a_i$ is \begin{equation}
P = \frac{{\bf F} \cdot \hat{\bf z}}{A} 
= \frac{1}{\text{i}} \int_{-\infty}^{\infty} \frac{d\zeta}{2\pi}
\int \frac{d^2k_\perp}{(2\pi)^2}\left[ T_{33}(a_i+\delta) -T_{33}(a_i-\delta)\right], 
\label{P-ai}
\end{equation} where \begin{equation}
    T_{33}(z) = \frac{1}{2}\Big[E_1^2 +E_2^2 -E_3^2\Big]\bigg|_{z} 
+ \frac{1}{2}\Big[H_1^2 +H_2^2 -H_3^2 \Big] \bigg|_{z}
\end{equation} from Eq. (\ref{stten-def}). Using this expression initially we are interested in measuring the force between two plates $\bf F^{\text{2-$\delta$}}$ (Superscript 2-$\delta$ refers to force on two plates), using the Green's function of two plates in regions of Fig. 1(b). The force on plate at $z=a_2$ is derived from 

 \begin{equation}
 P^{\text{2-$\delta$}}|_{z=a_2} = \frac{{\bf F^{\text{2-$\delta$}}} \cdot \hat{\bf z}}{A} \bigg|_{z=a_2}
 = \frac{1}{\text{i}} \int_{-\infty}^{\infty} \frac{d\zeta}{2\pi}
 \int \frac{d^2k_\perp}{(2\pi)^2}\left[ T_{33}(a_2+\delta) -T_{33}(a_2-\delta)\right],
 \end{equation}
 where
 \begin{multline}
     T_{33}(a_2+\delta) = \frac{1}{2\text{i}}\Big[\partial \partial'g^{2,H}_{ \rput(0.1,0){\scriptstyle{13}}
 \pscircle[linewidth=0.2pt](0.1,0.01){0.15} }
 (a_2,a_2)-\zeta^2 g^{2,E}_{ \rput(0.1,0){\scriptstyle{13}}
 \pscircle[linewidth=0.2pt](0.1,0.01){0.15} }
 (a_2,a_2) -k_{\perp}^2 g^{2,H}_{ \rput(0.1,0){\scriptstyle{13}}
 \pscircle[linewidth=0.2pt](0.1,0.01){0.15} }
 (a_2,a_2)\Big] \\ +\frac{1}{2\text{i}}\Big[\partial \partial'g^{2,E}_{ \rput(0.1,0){\scriptstyle{13}}
 \pscircle[linewidth=0.2pt](0.1,0.01){0.15} }
 (a_2,a_2)-\zeta^2 g^{2,H}_{ \rput(0.1,0){\scriptstyle{13}}
 \pscircle[linewidth=0.2pt](0.1,0.01){0.15} }
 (a_2,a_2) -k_{\perp}^2 g^{2,E}_{ \rput(0.1,0){\scriptstyle{13}}
 \pscircle[linewidth=0.2pt](0.1,0.01){0.15} }
 (a_2,a_2) \Big] 
 \end{multline}
 and
 \begin{multline}
 T_{33}(a_2-\delta) = \frac{1}{2\text{i}}\Big[\partial \partial'g^{2,H}_{ \rput(0.1,0){\scriptstyle{22}}
 \pscircle[linewidth=0.2pt](0.1,0.01){0.15} }
 (a_2,a_2)-\zeta^2 g^{2,E}_{ \rput(0.1,0){\scriptstyle{22}}
 \pscircle[linewidth=0.2pt](0.1,0.01){0.15} }
 (a_2,a_2) -k_{\perp}^2 g^{2,H}_{ \rput(0.1,0){\scriptstyle{22}}
 \pscircle[linewidth=0.2pt](0.1,0.01){0.15} }
 (a_2,a_2)\Big] \\ +\frac{1}{2\text{i}}\Big[\partial \partial'g^{2,E}_{ \rput(0.1,0){\scriptstyle{22}}
 \pscircle[linewidth=0.2pt](0.1,0.01){0.15} }
 (a_2,a_2)-\zeta^2 g^{2,H}_{ \rput(0.1,0){\scriptstyle{22}}
 \pscircle[linewidth=0.2pt](0.1,0.01){0.15} }
 (a_2,a_2) -k_{\perp}^2 g^{2,E}_{ \rput(0.1,0){\scriptstyle{22}}
 \pscircle[linewidth=0.2pt](0.1,0.01){0.15} }
 (a_2,a_2) \Big].   \end{multline}
 Implementing two plate Green's function force between the two plates can be obtained as \begin{equation}
 P^{\text{2-$\delta$}}|_{z=a_2} = \frac{{\bf F^{\text{2-$\delta$}}} \cdot \hat{\bf z}}{A} \bigg|_{z=a_2}
 = -\frac{1}{2 \pi^2} \int_{0}^{\infty} \kappa^3 d\kappa \Bigg[\frac{r_1^H r_2^H \hbox{e}^{-2\kappa a}}{\Delta_{12}^H}+\frac{r_1^E r_2^E \text{e}^{-2\kappa a}}{\Delta_{12}^E}\Bigg]. \label{F2}
 \end{equation}
 Similarly, the force between three plates $\bf F^{\text{3-$\delta$}}$ is obtained by involving the Green's function of three plates in regions of Fig. 2. The calculation of force on the plate at $z=a_3$ from Eq. (\ref{P-ai}) involves \begin{multline}
     T_{33}(a_3+\delta) = \frac{1}{2\text{i}}\Big[\partial \partial'g^{3,H}_{ \rput(0.1,0){\scriptstyle{14}}
 \pscircle[linewidth=0.2pt](0.1,0.01){0.15} }
 (a_3,a_3)-\zeta^2 g^{3,E}_{ \rput(0.1,0){\scriptstyle{14}}
 \pscircle[linewidth=0.2pt](0.1,0.01){0.15} }
 (a_3,a_3) -k_{\perp}^2 g^{3,H}_{ \rput(0.1,0){\scriptstyle{14}}
 \pscircle[linewidth=0.2pt](0.1,0.01){0.15} }
 (a_3,a_3)\Big] \\ +\frac{1}{2\text{i}}\Big[\partial \partial'g^{3,E}_{ \rput(0.1,0){\scriptstyle{14}}
 \pscircle[linewidth=0.2pt](0.1,0.01){0.15} }
 (a_3,a_3)-\zeta^2 g^{3,H}_{ \rput(0.1,0){\scriptstyle{14}}
 \pscircle[linewidth=0.2pt](0.1,0.01){0.15} }
 (a_3,a_3) -k_{\perp}^2 g^{3,E}_{ \rput(0.1,0){\scriptstyle{14}}
 \pscircle[linewidth=0.2pt](0.1,0.01){0.15} }
 (a_3,a_3) \Big] 
 \end{multline}
 and
 \begin{multline}
 T_{33}(a_3-\delta) = \frac{1}{2\text{i}}\Big[\partial \partial'g^{3,H}_{ \rput(0.1,0){\scriptstyle{23}}
 \pscircle[linewidth=0.2pt](0.1,0.01){0.15} }
 (a_3,a_3)-\zeta^2 g^{3,E}_{ \rput(0.1,0){\scriptstyle{23}}
 \pscircle[linewidth=0.2pt](0.1,0.01){0.15} }
 (a_3,a_3) -k_{\perp}^2 g^{3,H}_{ \rput(0.1,0){\scriptstyle{23}}
 \pscircle[linewidth=0.2pt](0.1,0.01){0.15} }
 (a_3,a_3)\Big] \\ +\frac{1}{2\text{i}}\Big[\partial \partial'g^{3,E}_{ \rput(0.1,0){\scriptstyle{23}}
 \pscircle[linewidth=0.2pt](0.1,0.01){0.15} }
 (a_3,a_3)-\zeta^2 g^{3,H}_{ \rput(0.1,0){\scriptstyle{23}}
 \pscircle[linewidth=0.2pt](0.1,0.01){0.15} }
 (a_3,a_3) -k_{\perp}^2 g^{3,E}_{ \rput(0.1,0){\scriptstyle{23}}
 \pscircle[linewidth=0.2pt](0.1,0.01){0.15} }
 (a_3,a_3) \Big].   \end{multline} Further, the force between the three plates is evaluated as 
 \begin{multline}
 P^{\text{3-$\delta$}}|_{z=a_3} = \frac{{\bf F^{\text{3-$\delta$}}} \cdot \hat{\bf z}}{A} \bigg|_{z=a_3}
 = \\ -\frac{1}{2 \pi^2} \int_{0}^{\infty} \kappa^3 d\kappa \Bigg[\frac{r_2^H r_3^H \hbox{e}^{-2\kappa b} \Delta_{12}^H}{\Delta_{123}^H}+\frac{r_2^E r_3^E \text{e}^{-2\kappa b} \Delta_{12}^E }{\Delta_{123}^E}+\frac{r_1^H (t_2^H)^2 r_3^H \hbox{e}^{-2\kappa a} \hbox{e}^{-2\kappa b}}{\Delta_{123}^H}+\frac{r_1^E (t_2^E)^2 r_3^E \hbox{e}^{-2\kappa a} \hbox{e}^{-2\kappa b}}{\Delta_{123}^E}\Bigg]. \label{F3}
 \end{multline}
\section{CASIMIR ENERGY FROM MULTIPLE SCATTERING FORMALISM}
Here we are interested in calculating the Casimir energy between two plates ($\Delta E_{(12)}$), three plates ($\Delta E_{(123)}$), four plates ($\Delta E_{(1234)}$) and five plates ($\Delta E_{(12345)}$) configurations from multiple scattering formalism. From the Green's functions in Sec. \rom{3}, the reflection coefficients of one plate, two plates and three plates configurations are used to calculate these energies of multiple plates. The total energy $E_{(12\cdots i,i+1 \cdots N)}$ between two disjoint bodies $(12\cdots i)$ and $(i+1 \cdots N)$ can be decomposed as\,\cite{Shajesh2011} \begin{equation}
    E_{(12\cdots i,i+1 \cdots N)}=E_0+\Delta E_{(12\cdots i)}+\Delta E_{(i+1 \cdots N)}+\Delta E_{(12\cdots i)(i+1 \cdots N)}. \label{E2}
\end{equation} Here $E_0$ is the energy of the background vacuum which is usually infinite, $\Delta E_{(12\cdots i)}$, $\Delta E_{(i+1 \cdots N)}$ are self-energies of the bodies, and $\Delta E_{(12\cdots i)(i+1 \cdots N)}$ is the interaction energy. Self-energy can be interpreted as the energy needed to create the bodies independently. The ``finite part" of this total energy $E_{(12\cdots i,i+1 \cdots N)}$ after removing the divergent parts ($E_0$ and $\Delta E_i$, $i\in[1,N]$) is what we are interested in calculating $\Delta E_{(12\cdots i,i+1 \cdots N)}$, which is responsible for the Casimir force and the Casimir interaction energy $\Delta E_{(12\cdots i)(i+1 \cdots N)}$ is usually finite.

Multiple scattering formalism predicts this Casimir interaction energy $\Delta E_{(12\cdots i)(i+1 \cdots N)}$ between the two disjoint planar bodies $(12\cdots i)$ and $(i+1 \cdots N)$ with reflection coefficients $R_{(12\cdots i)}$ and $R_{(i+1 \cdots N)}$ (with superscript $H$ denoting TM mode and $E$ for TE mode, subscript $(12\cdots i)$ and $(i+1 \cdots N)$ denoting the optical properties of the two bodies), respectively as
\begin{equation}
 \frac{\Delta E_{(12\cdots i)(i+1 \cdots N)}}{A} =  \frac{1}{2} \int_{-\infty}^\infty \frac{d\zeta}{2\pi}
\int \frac{d^2k_\perp}{(2\pi)^2} \Bigg[ \ln \Big[ 1 - R_{(12\cdots i)}^{<,H} \text{e}^{-\kappa L} R_{(i+1 \cdots N)}^{>,H} \text{e}^{-\kappa L}\Big] + \ln \Big[ 1 - R_{(12\cdots i)}^{<,E} \text{e}^{-\kappa L} R_{(i+1 \cdots N)}^{>,E} \text{e}^{-\kappa L}\Big] \Bigg]
\end{equation}
where $L$ is the distance between the bodies. This assumes that Casimir interaction energy can be represented only using reflection coefficients alone. 

The Casimir energy between two plates $\Delta E_{(12)}$ with reflection coefficients $R_{(1)}=r_1$, $R_{(2)}=r_2$ (Eq. (\ref{coeff}) for $i=1,2$) and distance $l_{12}$ between them, can be easily evaluated as \begin{equation}
 \frac{\Delta E_{(12)}}{A} =  \frac{1}{2} \int_{-\infty}^\infty \frac{d\zeta}{2\pi}
\int \frac{d^2k_\perp}{(2\pi)^2} \Bigg[ \ln \Big[ \Delta_{12}^H \Big] + \ln \Big[ \Delta_{12}^E \Big] \Bigg],
\end{equation}
where $\Delta_{12}$ is multiple scattering parameter of two plates in Eq. (\ref{MS1}) such as \begin{equation}
    \Delta_{12}=1-r_1 \hbox{e}^{-\kappa l_{12}} r_2 \hbox{e}^{-\kappa l_{12}}.
    \label{MS12}
 \end{equation} In this case of two plate, $\Delta E_{(12)}$ = $\Delta E_{(1)(2)}$ since self-energies of plates $\Delta E_{(1)}$ and $\Delta E_{(2)}$ are infinite. This expression for energy can be checked by finding the force and comparing with $\bf F^{\text{2-$\delta$}}$ between the two plates (Eq. (\ref{F2})) where \begin{equation}
    P^{\text{2-$\delta$}}|_{z=a_2} = -\frac{1}{A}\frac{\partial \Delta E_{(12)}}{\partial a}
\end{equation} with $a=l_{12}$. 

The Casimir energy between three plates $\Delta E_{(123)}$ with reflection coefficients $R_{(1)}=r_1$, $R_{(2)}=r_2$, $R_{(3)}=r_3$ (Eq. (\ref{coeff}) for $i=1,2,3$) and with distances $l_{12}$, $l_{23}$ between them, respectively can be evaluated from using Eq. (\ref{E2}) such as 
\begin{equation}
        E_{(123)}=E_0+\Delta E_{(12)}+\Delta E_{(3)}+\Delta E_{(12)(3)}=E_0+\Delta E_{(1)}+ \Delta E_{(23)}+\Delta E_{(1)(23)}.
\end{equation}
Considering the finite terms in this expression, energy can be solved as 
\begin{multline}
       \frac{\Delta E_{(123)}}{A}= \frac{1}{2} \int_{-\infty}^\infty \frac{d\zeta}{2\pi}
\int \frac{d^2k_\perp}{(2\pi)^2} \Bigg[ \ln \Big[ \Delta_{12}^H \Big] + \ln \Big[ \Delta_{12}^E \Big] +\ln \Big[ 1 - R_{12}^{<,H} R_{3}^{>,H} \text{e}^{-2\kappa l_{23}}\Big] + \ln \Big[ 1 - R_{12}^{<,E} R_{3}^{>,E} \text{e}^{-2\kappa l_{23}}\Big]\Bigg]\\=\frac{1}{2} \int_{-\infty}^\infty \frac{d\zeta}{2\pi}
\int \frac{d^2k_\perp}{(2\pi)^2} \Bigg[ \ln \Big[ \Delta_{23}^H \Big] + \ln \Big[ \Delta_{23}^E \Big] +\ln \Big[ 1 - R_{1}^{<,H} R_{23}^{>,H} \text{e}^{-2\kappa l_{12}}\Big] + \ln \Big[ 1 - R_{1}^{<,E} R_{23}^{>,E} \text{e}^{-2\kappa l_{12}}\Big]\Bigg].
\end{multline}
Reflection coefficients $R_{12}^{<}$ and $R_{23}^{>}$ can be obtained from matrix $B$ of two plates Green's function. The Casimir energy for three plates is derived as \begin{equation}
 \frac{\Delta E_{(123)}}{A} =  \frac{1}{2} \int_{-\infty}^\infty \frac{d\zeta}{2\pi}
\int \frac{d^2k_\perp}{(2\pi)^2} \Bigg[ \ln \Big[ \Delta_{123}^H \Big] + \ln \Big[ \Delta_{123}^E \Big] \Bigg],
\end{equation}
where $\Delta_{123}$ is multiple scattering parameter of three plates in Eq. (\ref{MS3}) which can be defined as Eq. (\ref{MS123}) with
\begin{equation}
    \Delta_{13}=-r_1^H \hbox{e}^{-\kappa l_{12}}  t_2^H \hbox{e}^{-\kappa l_{23}} r_3^H \hbox{e}^{-\kappa l_{23}} t_2^H \hbox{e}^{-\kappa l_{12}} \label{MS13}
\end{equation} and using this we define the generalized multiple scattering parameter such as \begin{equation}
    \Delta_{ij}= 1-r_i \hbox{e}^{-\kappa l_{ij}} r_j \hbox{e}^{-\kappa l_{ij}} \label{nn}
\end{equation} for all $j=i+1$ (nearest neighbour scattering, $i\in[1,N-1]$ where $i$ and $j$ are adjacent). These terms in $\Delta_{123}$ can be visualized in Fig. \ref{Del123} where $\Delta_{12}$ and $\Delta_{23}$ are loops between bodies $(1)-(2)$ and $(2)-(3)$, respectively. Term $\Delta_{13}$ in Eq.(\ref{MS13}) refers to loop between $(1)-(3)$ with initial reflection with $r_1$, propagation with exponential dependence of length $l_{12}$, transmission with $t_2$, propagation of length $l_{23}$, reflection with $r_3$, again propagation of length $l_{23}$, transmission with $t_2$, propagation of length $l_{12}$ and this propagation continues. Further, these loops representation can be seen for multiple scattering parameters $\Delta_{1234}$, $\Delta_{12345}$ of $N=4,5$ configurations. Similarly, this expression of energy for three plates can be checked by finding the force and comparing it with $\bf F^{\text{3-$\delta$}}$ between the three plates (Eq. (\ref{F3})) where \begin{equation}
    P^{\text{3-$\delta$}}|_{z=a_3} = -\frac{1}{A}\frac{\partial \Delta E_{(123)}}{\partial b}.
\end{equation} with $a=l_{12}$ and $b=l_{23}$. 

In a similar manner, Casimir energy between four plates $\Delta E_{(1234)}$ with reflection coefficients $R_{(1)}=r_1$, $R_{(2)}=r_2$, $R_{(3)}=r_3$, $R_{(4)}=r_4$ (Eq. (\ref{coeff}) for $i=1,2,3,4$) and with distances $l_{12}$, $l_{23}$, $l_{34}$ between them, respectively can be evaluated as \begin{equation}
 \frac{\Delta E_{(1234)}}{A} =  \frac{1}{2} \int_{-\infty}^\infty \frac{d\zeta}{2\pi}
\int \frac{d^2k_\perp}{(2\pi)^2} \Bigg[ \ln \Big[ \Delta_{1234}^H \Big] + \ln \Big[ \Delta_{1234}^E \Big] \Bigg],
\end{equation} 
where multiple scattering parameter of four plates is defined in Eq. (\ref{MS1234}). Here $\Delta_{ij}$ for $j=i+1$ is defined in Eq. (\ref{nn}), $\Delta_{13}$ is same as in Eq.(\ref{MS123}), \begin{equation}
    \Delta_{24}=-r_2 \hbox{e}^{-\kappa l_{23}}  t_3 \hbox{e}^{-\kappa l_{34}} r_4 \hbox{e}^{-\kappa l_{34}} t_3 \hbox{e}^{-\kappa l_{23}}\end{equation} and
 \begin{equation}
     \Delta_{14}=-r_1 \hbox{e}^{-\kappa l_{12}}  t_2 \hbox{e}^{-\kappa l_{23}} t_3 \hbox{e}^{-\kappa l_{34}} r_4 \hbox{e}^{-\kappa l_{34}} t_3 \hbox{e}^{-\kappa l_{23}} t_2 \hbox{e}^{-\kappa l_{12}}. \end{equation} Using this we define \begin{equation}
    \Delta_{ik}= -r_i \hbox{e}^{-\kappa l_{i,i+1}} t_{i+1} \hbox{e}^{-\kappa l_{i+1,i+2}} t_{i+2} \cdots \hbox{e}^{-\kappa l_{k-1,k}} r_k \hbox{e}^{-\kappa l_{k-1,k}} \cdots t_{i+1} \hbox{e}^{-\kappa l_{i,i+1}}
\end{equation} for all $k\geq i+2$ (next-to-nearest neighbour, next-to-next-to-nearest neighbour, $\cdots$ scattering, $i\in[1,N-2]$ where $i$ and $k$ are not adjacent). This Casimir energy between four plates $\Delta E_{(1234)}$ was evaluated from using Eq. (\ref{E2}) such as 
\begin{equation}
         E_{(1234)}=E_0+\Delta E_{(12)}+\Delta E_{(34)}+\Delta E_{(12)(34)}=E_0+\Delta E_{(1)}+ \Delta E_{(234)}+\Delta E_{(1)(234)}=E_0+\Delta E_{(123)}+ \Delta E_{(4)}+\Delta E_{(123)(4)}.
\end{equation} We compared the Casimir energy obtained for three plates from multiple scattering formalism with Casimir force obtained from the stress tensor method in Eq. (70). This ensures the multiple scattering formalism for disjoint many bodies we have implemented and also checks the reflection coefficients ($R_{12}$ and $R_{23}$ in Eq. (66)) obtained for two plates configuration. From different ways of obtaining Casimir energy for four plates in Eq. (75), we check the reflection coefficients obtained for three plates configuration ($R_{234}$ and $R_{123}$ in $\Delta E_{(1)(234)}$ and $\Delta E_{(123)(4)}$, respectively) implementing only reflection coefficients for two plates configuration ($R_{12}$ and $R_{23}$ in $\Delta E_{(12)(34)}$). For five plates the multiple scattering parameter can be obtained, such as \begin{equation}
\Delta_{12345}=\Delta_{12}\Delta_{23}\Delta_{34}\Delta_{45}+\Delta_{12}\Delta_{24}\Delta_{45}+\Delta_{13}\Delta_{34}\Delta_{45}+\Delta_{12}\Delta_{23}\Delta_{35}+\Delta_{13}\Delta_{35}+\Delta_{12}\Delta_{25}+\Delta_{14}\Delta_{45}+\Delta_{15}, \label{MS12345}
\end{equation} where the Casimir energy between five plates $\Delta E_{(12345)}$ was evaluated from using Eq. (\ref{E2}) such as 
\begin{equation}
          E_{(12345)}=E_0+\Delta E_{(123)}+\Delta E_{(45)}+\Delta E_{(123)(45)}=E_0+\Delta E_{(12)}+ \Delta E_{(345)}+\Delta E_{(12)(345)}.
\end{equation} These multiple scattering parameters can be visualized using a simple diagrammatic way, such as in Figs. \ref{Del1234} and 5 with each term described by loops. These results for $\Delta$ are similar to the expressions obtained by Allocca et al.\,\cite{Allocca2022} for $N=2,3,4,5$ (Multiple scattering parameter $\Delta_{12}$, $\Delta_{123}$, $\Delta_{1234}$, $\Delta_{12345}$ here referred to as generating function $\Delta_1$, $\Delta_2$, $\Delta_3$, $\Delta_4$, respectively in Ref.\,\cite{Allocca2022,Rosa2017}). Further, using this similar approach of separating partitions such as in Eq.s (\ref{MS123}), (\ref{MS1234}) and (\ref{MS12345}), Casimir energy of N $\delta$-plates can be generalized such as Eq. (1) where $\Delta_{12\cdots N}$ can be evaluated from the discussion above.
  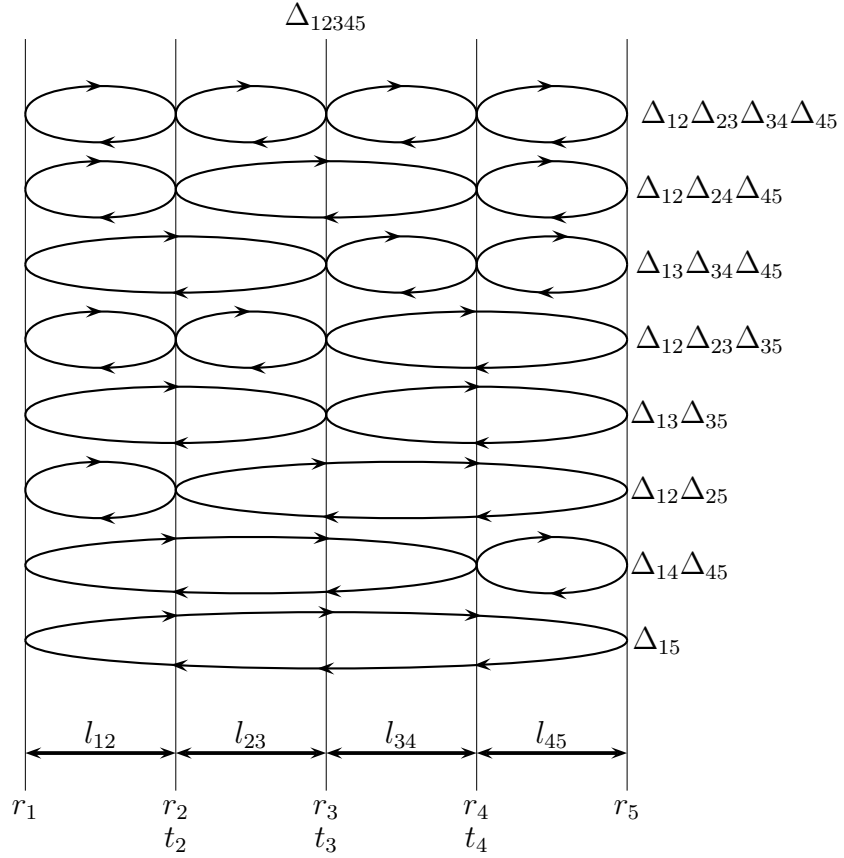
\begin{figure}
    \centering 
\begin{pspicture}(8,10)(0,0)
  \psline[linewidth=0mm]{-}(0,0)(0,10)
  \psline[linewidth=0mm]{-}(2,0)(2,10)
  \psline[linewidth=0mm]{-}(4,0)(4,10)
  \psline[linewidth=0mm]{-}(6,0)(6,10)
  \psline[linewidth=0mm]{-}(8,0)(8,10)
  \psline[linewidth=0.5mm]{<->}(0,0.5)(2,0.5)
  \rput(1,0.75){\textcolor{black}{ \large $l_{12}$}}
\psline[linewidth=0.5mm]{<->}(2,0.5)(4,0.5)
  \rput(3,0.75){\textcolor{black}{ \large $l_{23}$}}
  \psline[linewidth=0.5mm]{<->}(4,0.5)(6,0.5)
  \rput(5,0.75){\textcolor{black}{ \large $l_{34}$}}
  \psline[linewidth=0.5mm]{<->}(6,0.5)(8,0.5)
  \rput(7,0.75){\textcolor{black}{ \large $l_{45}$}}
\psset{arrowscale=1.5}

\psbezier[ArrowInside=->,showpoints=false,ArrowInsideNo=1](0,9)(0,9.5)(2,9.5)(2,9)
 \psbezier[ArrowInside=->,showpoints=false,ArrowInsideNo=1](2,9)(2,8.5)(0,8.5)(0,9)

 \psbezier[ArrowInside=->,showpoints=false,ArrowInsideNo=1](2,9)(2,9.5)(4,9.5)(4,9)
 \psbezier[ArrowInside=->,showpoints=false,ArrowInsideNo=1](4,9)(4,8.5)(2,8.5)(2,9)

 \psbezier[ArrowInside=->,showpoints=false,ArrowInsideNo=1](4,9)(4,9.5)(6,9.5)(6,9)
 \psbezier[ArrowInside=->,showpoints=false,ArrowInsideNo=1](6,9)(6,8.5)(4,8.5)(4,9)
\psbezier[ArrowInside=->,showpoints=false,ArrowInsideNo=1](6,9)(6,9.5)(8,9.5)(8,9)
 \psbezier[ArrowInside=->,showpoints=false,ArrowInsideNo=1](8,9)(8,8.5)(6,8.5)(6,9)
 \psbezier[ArrowInside=->,showpoints=false,ArrowInsideNo=1](0,8)(0,8.5)(2,8.5)(2,8)
 \psbezier[ArrowInside=->,showpoints=false,ArrowInsideNo=1](2,8)(2,7.5)(0,7.5)(0,8)
 \psbezier[ArrowInside=->,showpoints=false,ArrowInsideNo=1](2,8)(2,8.5)(6,8.5)(6,8)
 \psbezier[ArrowInside=->,showpoints=false,ArrowInsideNo=1](6,8)(6,7.5)(2,7.5)(2,8)

 \psbezier[ArrowInside=->,showpoints=false,ArrowInsideNo=1](6,8)(6,8.5)(8,8.5)(8,8)
 \psbezier[ArrowInside=->,showpoints=false,ArrowInsideNo=1](8,8)(8,7.5)(6,7.5)(6,8)
 \psbezier[ArrowInside=->,showpoints=false,ArrowInsideNo=1](0,7)(0,7.5)(4,7.5)(4,7)
 \psbezier[ArrowInside=->,showpoints=false,ArrowInsideNo=1](4,7)(4,6.5)(0,6.5)(0,7)
 \psbezier[ArrowInside=->,showpoints=false,ArrowInsideNo=1](4,7)(4,7.5)(6,7.5)(6,7)
 \psbezier[ArrowInside=->,showpoints=false,ArrowInsideNo=1](6,7)(6,6.5)(4,6.5)(4,7)
\psbezier[ArrowInside=-<,showpoints=false,ArrowInsideNo=1](6,7)(6,6.5)(8,6.5)(8,7)
 \psbezier[ArrowInside=-<,showpoints=false,ArrowInsideNo=1](8,7)(8,7.5)(6,7.5)(6,7)
 \psbezier[ArrowInside=->,showpoints=false,ArrowInsideNo=1](0,6)(0,6.5)(2,6.5)(2,6)
 \psbezier[ArrowInside=->,showpoints=false,ArrowInsideNo=1](2,6)(2,5.5)(0,5.5)(0,6)

\psbezier[ArrowInside=->,showpoints=false,ArrowInsideNo=1](2,6)(2,6.5)(4,6.5)(4,6)
 \psbezier[ArrowInside=->,showpoints=false,ArrowInsideNo=1](4,6)(4,5.5)(2,5.5)(2,6)

\psbezier[ArrowInside=->,showpoints=false,ArrowInsideNo=1](4,6)(4,6.5)(8,6.5)(8,6)
 \psbezier[ArrowInside=->,showpoints=false,ArrowInsideNo=1](8,6)(8,5.5)(4,5.5)(4,6)

 \psbezier[ArrowInside=->,showpoints=false,ArrowInsideNo=1](0,5)(0,5.5)(4,5.5)(4,5)
 \psbezier[ArrowInside=->,showpoints=false,ArrowInsideNo=1](4,5)(4,4.5)(0,4.5)(0,5)


 \psbezier[ArrowInside=->,showpoints=false,ArrowInsideNo=1](4,5)(4,5.5)(8,5.5)(8,5)
 \psbezier[ArrowInside=->,showpoints=false,ArrowInsideNo=1](8,5)(8,4.5)(4,4.5)(4,5)


\psbezier[ArrowInside=->,showpoints=false,ArrowInsideNo=1](0,4)(0,4.5)(2,4.5)(2,4)
 \psbezier[ArrowInside=->,showpoints=false,ArrowInsideNo=1](2,4)(2,3.5)(0,3.5)(0,4)

\psbezier[ArrowInside=->,showpoints=false,ArrowInsideNo=2](2,4)(2,4.5)(8,4.5)(8,4)
 \psbezier[ArrowInside=->,showpoints=false,ArrowInsideNo=2](8,4)(8,3.5)(2,3.5)(2,4)

\psbezier[ArrowInside=->,showpoints=false,ArrowInsideNo=2](0,3)(0,3.5)(6,3.5)(6,3)
 \psbezier[ArrowInside=->,showpoints=false,ArrowInsideNo=2](6,3)(6,2.5)(0,2.5)(0,3)

\psbezier[ArrowInside=->,showpoints=false,ArrowInsideNo=1](6,3)(6,3.5)(8,3.5)(8,3)
 \psbezier[ArrowInside=->,showpoints=false,ArrowInsideNo=1](8,3)(8,2.5)(6,2.5)(6,3)

 \psbezier[ArrowInside=->,showpoints=false,ArrowInsideNo=3](0,2)(0,2.5)(8,2.5)(8,2)
\psbezier[ArrowInside=->,showpoints=false,ArrowInsideNo=3](8,2)(8,1.5)(0,1.5)(0,2)

\rput(0,-0.25){\textcolor{black}{ \large $r_1$}}
\rput(2,-0.25){\textcolor{black}{ \large $r_2$}}
\rput(2,-0.65){\textcolor{black}{ \large $t_2$}}
\rput(4,-0.25){\textcolor{black}{ \large $r_3$}}
\rput(4,-0.65){\textcolor{black}{ \large $t_3$}}
\rput(6,-0.25){\textcolor{black}{ \large $r_4$}}
\rput(6,-0.65){\textcolor{black}{ \large $t_4$}}
\rput(8,-0.25){\textcolor{black}{ \large $r_5$}}
\rput(4,10.3){\textcolor{black}{ \large $\Delta_{12345}$}}
\rput(9.5,9){\textcolor{black}{ \large $\Delta_{12} \Delta_{23}\Delta_{34}\Delta_{45}$}}
\rput(9.1,8){\textcolor{black}{ \large $\Delta_{12} \Delta_{24}\Delta_{45}$}}
\rput(9.1,7){\textcolor{black}{ \large $\Delta_{13} \Delta_{34}\Delta_{45}$}}
\rput(9.1,6){\textcolor{black}{ \large $\Delta_{12} \Delta_{23}\Delta_{35}$}}
\rput(8.7,5){\textcolor{black}{ \large $\Delta_{13} \Delta_{35}$}}
\rput(8.7,4){\textcolor{black}{ \large $\Delta_{12} \Delta_{25}$}}
\rput(8.7,3){\textcolor{black}{ \large $\Delta_{14} \Delta_{45}$}}
\rput(8.4,2){\textcolor{black}{ \large $\Delta_{15}$}}
\end{pspicture}
\vspace{6mm}
    \caption{Visualizing $\Delta_{12345}$ with optical properties, $r_i$ are reflection coefficients and $t_i$ are transmission coefficients of the plates. $l_{ij}$ is the distance between the plates where $i$ and $j$ are adjacent plates.}
\end{figure}
\subsection{Perfectly dielectric $N$ $\delta$-function plates}

For analysis of our results, we consider a simple case of $N$ perfect electrically conducting ($\lambda_{ei} \rightarrow \infty$, $\lambda_{gi}=0$) $\delta$-function plates with distances $l_{ij}$ between the nearest plates $\{i,j\}$ and optical properties $r^H_i=1,r^E_i=-1,t^H_i=t^E_i=0$. In this case, the Casimir energy for $N$ $\delta-$function plates is obtained from Eq. (1) as 
\begin{multline}
 \frac{\Delta E_{(12\cdots N)}}{A} = \frac{1}{2} \int_{-\infty}^\infty \frac{d\zeta}{2\pi}
\int \frac{d^2k_\perp}{(2\pi)^2} \Bigg[ \ln \Big[ \Delta_{12}^H \Delta_{23}^H \cdots \Delta_{N-1,N}^H\Big] + \ln \Big[ \Delta_{12}^E \Delta_{23}^E \cdots \Delta_{N-1,N}^E\Big] \Bigg] \\= \frac{1}{2} \int_{-\infty}^\infty \frac{d\zeta}{2\pi}
\int \frac{d^2k_\perp}{(2\pi)^2} 2 \ln \Big[ (1
- e^{-2\kappa l_{12}})(1
- e^{-2\kappa l_{23}})\cdots(1
- e^{-2\kappa l_{N-1,N}})\Big] = -\frac{\pi^2}{720}\sum_{i=1}^{N-1} \frac{1}{l_{i,i+1}^3}.
\end{multline}
This result is straightforward since only the nearest neighbour scattering $\Delta_{ij}$ is present. For the more specific case of uniform distances between the perfectly conducting dielectric plates $l_{ij}=d$, the Casimir energy of $N$ plates is $\Delta E_{(12\cdots N)}=(N-1)\Delta E_{(12)}$, where Casimir energy between nearest two plates is $\Delta E_{(12)}/A = -\pi^2/720d^3$ \cite{Casimir:1948dh} and consecutively, the force is $\bf F^{\text{2-$\delta$}}\cdot \hat{\bf z}$$/A=\pi^2/240d^4$ \cite{Lifshitz:1956zz} from Eq. (64). This shows that increasing the dielectric plates leads to intensifying of the Casimir force as demonstrated by Allocca et al. \cite{Allocca2022}. The result in Eq. (78) was shown previously for scalar case by considering $N$ $\delta$-function plates satisfying Dirichlet boundary conditions \cite{selfsimilar}.
\section{Conclusions}
The Casimir energy for $N=2,3,4,5$ $\delta-$function plates was derived using the multiple scattering formalism and by implementing the optical properties of plates. The pattern seems generic, and the result for $N$ plates was predicted by extrapolation and organising Casimir energy density derived from multiple scattering parameter $\Delta$ into nearest neighbour scattering and next-to-nearest neighbour scattering terms. Visualization of the distribution of scattering terms in multiple scattering parameter using diagrammatic loops appears to be an easy way to obtain $N$ plates result. 

Green's functions were derived for two and three $\delta-$function plates configurations, using which their reflection coefficients were obtained. Multiple scattering formalism was utilised to derive the Casimir energy between two, three, four and five plates configurations. In the case of Casimir energy for two plates configuration, reflection coefficients of a single plate were used. In the case of Casimir energy for three plates, reflection coefficients of a single plate and two plates configurations were used. In the case of Casimir energy for four plates, reflection coefficients of a single plate and three plates configurations were used. Similarly, in the case of Casimir energy for five plates, reflection coefficients of two plates and three plates configurations were used. For the case of two and three plates configurations, the Casimir energy was checked with force derived from the stress tensor method. Further, one can obtain an exact closed-form expression for $N$ plates. It may also be interesting to expand the matrices defined in Sec. \rom{3} in order to derive Green's functions for $N$ plates in a systematic manner. Investigating the Casimir energy for three plates configuration may also have relevance in the context of Casimir pistons\,\cite{cp1,cp2,cp3,cp4}.

 \section*{Acknowledgements}
 We are incredibly grateful to Dr. K. V. Shajesh and Dr. Prachi Parashar for introducing their work on $\delta$-function plates and for guidance throughout this work. We thank Dr. R. Sankaranarayanan at NITT for constant support and making all this happen in the beginning.   
\bibliographystyle{ieeetr}
\bibliography{sample.bib}

\begin{thebibliography}{10}

\bibitem{Casimir:1948dh}
H.~B.~G. Casimir, ``{On the Attraction Between Two Perfectly Conducting
  Plates},'' {\em Kon. Ned. Akad. Wetensch. Proc.}, vol.~51, p.~793, 1948.

\bibitem{Lifshitz:1956zz}
E.~M. Lifshitz, ``{The theory of molecular attractive forces between solids},''
  {\em Sov. Phys. JETP}, vol.~2, pp.~73--83, 1956.

\bibitem{Milton_2008}
K.~A. Milton and J.~Wagner, ``Multiple scattering methods in casimir
  calculations,'' {\em Journal of Physics A: Mathematical and Theoretical},
  vol.~41, p.~155402, apr 2008.

\bibitem{mult3}
M.~S. Toma\ifmmode~\check{s}\else \v{s}\fi{}, ``{Casimir} force in absorbing
  multilayers,'' {\em Phys. Rev. A}, vol.~66, p.~052103, 2002.

\bibitem{mult4}
M.~S. Toma\ifmmode~\check{s}\else \v{s}\fi{}, ``Casimir effect across a layered
  medium,'' {\em International Journal of Modern Physics: Conference Series},
  vol.~14, pp.~561--565, 2012.

\bibitem{Allocca2022}
A.~Allocca, S.~Avino, S.~Balestrieri, E.~Calloni, S.~Caprara, M.~Carpinelli,
  L.~D'Onofrio, D.~D'Urso, R.~De~Rosa, L.~Errico, G.~Gagliardi, M.~Grilli,
  V.~Mangano, M.~Marsella, L.~Naticchioni, A.~Pasqualetti, G.~P. Pepe,
  M.~Perciballi, L.~Pesenti, P.~Puppo, P.~Rapagnani, F.~Ricci, L.~Rosa,
  C.~Rovelli, D.~Rozza, P.~Ruggi, N.~Saini, V.~Sequino, V.~Sipala,
  D.~Stornaiuolo, F.~Tafuri, A.~Tagliacozzo, I.~Tosta~e Melo, and L.~Trozzo,
  ``{Casimir} energy for n superconducting cavities: a model for the {YBCO
  (GdBCO)} sample to be used in the archimedes experiment,'' {\em The European
  Physical Journal Plus}, vol.~137, no.~7, p.~826, 2022.

\bibitem{Barton_2005}
G.~Barton, ``{Casimir} effects for a flat plasma sheet: I. energies,'' {\em
  Journal of Physics A: Mathematical and General}, vol.~38, no.~13,
  pp.~2997--3019, 2005.

\bibitem{Prachi2012}
P.~Parashar, K.~A. Milton, K.~V. Shajesh, and M.~Schaden, ``Electromagnetic
  semitransparent $\ensuremath{\delta}$-function plate: {Casimir} interaction
  energy between parallel infinitesimally thin plates,'' {\em Phys. Rev. D},
  vol.~86, p.~085021, Oct 2012.

\bibitem{BALIAN1977}
R.~Balian and B.~Duplantier, ``Electromagnetic waves near perfect conductors.
  i. multiple scattering expansions. distribution of modes,'' {\em Annals of
  Physics}, vol.~104, no.~2, pp.~300--335, 1977.

\bibitem{BALIAN1978}
R.~Balian and B.~Duplantier, ``Electromagnetic waves near perfect conductors.
  ii. casimir effect,'' {\em Annals of Physics}, vol.~112, no.~1, pp.~165--208,
  1978.

\bibitem{Kenneth2006}
O.~Kenneth and I.~Klich, ``Opposites attract: A theorem about the {Casimir}
  force,'' {\em Phys. Rev. Lett.}, vol.~97, p.~160401, Oct 2006.

\bibitem{PRLemig}
T.~Emig, N.~Graham, R.~L. Jaffe, and M.~Kardar, ``Casimir forces between
  arbitrary compact objects,'' {\em Phys. Rev. Lett.}, vol.~99, p.~170403, Oct
  2007.

\bibitem{Milton2013}
K.~A. {Milton}, P.~{Parashar}, M.~{Schaden}, and K.~V. {Shajesh}, ``{Casimir
  interaction energies for magneto-electric $\delta$-function plates},'' {\em
  Nuovo Cimento C Geophysics Space Physics C}, vol.~36, pp.~193--204, May 2013.

\bibitem{Kenneth2008}
O.~Kenneth and I.~Klich, ``Casimir forces in a t-operator approach,'' {\em
  Phys. Rev. B}, vol.~78, p.~014103, Jul 2008.

\bibitem{SCHWINGER1978}
J.~Schwinger, L.~L. DeRaad, and K.~A. Milton, ``{Casimir} effect in
  dielectrics,'' {\em Annals of Physics}, vol.~115, no.~1, pp.~1--23, 1978.

\bibitem{Brevik2018}
I.~Brevik, P.~Parashar, and K.~V. Shajesh, ``{Casimir} force for
  magnetodielectric media,'' {\em Phys. Rev. A}, vol.~98, p.~032509, Sep 2018.

\bibitem{SHAJESH2017}
K.~Shajesh, P.~Parashar, and I.~Brevik, ``{Casimir–Polder} energy for axially
  symmetric systems,'' {\em Annals of Physics}, vol.~387, pp.~166--202, 2017.

\bibitem{schwinger1998classical}
J.~Schwinger, L.~Deraad, K.~Milton, W.~Tsai, and J.~Norton, {\em Classical
  Electrodynamics}.
\newblock Advanced book program, Avalon Publishing, 1998.

\bibitem{Shajesh2011}
K.~V. Shajesh and M.~Schaden, ``Many-body contributions to {Green's} functions
  and {Casimir} energies,'' {\em Phys. Rev. D}, vol.~83, p.~125032, Jun 2011.

\bibitem{Rosa2017}
L.~Rosa, S.~Avino, E.~Calloni, S.~Caprara, M.~De~Laurentis, R.~De~Rosa,
  G.~Esposito, M.~Grilli, E.~Majorana, G.~P. Pepe, S.~Petrarca, P.~Puppo,
  P.~Rapagnani, F.~Ricci, C.~Rovelli, P.~Ruggi, N.~L. Saini, C.~Stornaiolo, and
  F.~Tafuri, ``Casimir energy for two and three superconducting coupled
  cavities: Numerical calculations,'' {\em The European Physical Journal Plus},
  vol.~132, p.~478, Nov 2017.

\bibitem{selfsimilar}
K.~V. Shajesh, I.~Brevik, I.~Cavero-Pel\'aez, and P.~Parashar, ``Casimir
  energies of self-similar plate configurations,'' {\em Phys. Rev. D}, vol.~94,
  p.~065003, Sep 2016.

\bibitem{cp1}
R.~M. Cavalcanti, ``Casimir force on a piston,'' {\em Phys. Rev. D}, vol.~69,
  p.~065015, Mar 2004.

\bibitem{cp2}
E.~Elizalde, S.~D. Odintsov, and A.~A. Saharian, ``Repulsive casimir effect
  from extra dimensions and robin boundary conditions: From branes to
  pistons,'' {\em Phys. Rev. D}, vol.~79, p.~065023, Mar 2009.

\bibitem{cp3}
V.~K. Oikonomou, ``Casimir pistons with curved boundaries,'' {\em Modern
  Physics Letters A}, vol.~24, no.~30, pp.~2405--2423, 2009.

\bibitem{cp4}
V.~K. Oikonomou and N.~D. Tracas, ``Slab bag fermionic {C}asimir effect, chiral
  boundaries and vector boson-{M}ajorna fermion pistons,'' {\em International
  Journal of Modern Physics A}, vol.~25, no.~32, pp.~5935--5950, 2010.

\end{thebibliography}
\end{document}